\documentclass[twocolumn,amsmath,amssymb,prd,preprintnumbers,showpacs]{revtex4}

\usepackage{graphicx}
\usepackage{latexsym}
\usepackage{bm}

\newcommand{\beq}[1]{\begin{equation}\label{#1}}
\newcommand{\eeq}{\end{equation}}
\newcommand{\bea}[1]{\begin{eqnarray} \label{#1}}
\newcommand{\eea}{\end{eqnarray}}
\newcommand{\ba}{\begin{array}}
\newcommand{\ea}{\end{array}}

\newcommand{\rf}[1]{(\ref{#1})}

\newcommand{\half}{\frac{1}{2}}

\newcommand{\Wb}{W\!\!\cdot\!b}
\newcommand{\sdot}{\!\cdot\!}

\def\fr#1#2{{{#1} \over {#2}}}
\def\etal{{\it et al.}}

\begin{document}

\preprint{MIT-CTP-3771}

\title{Non-local on-shell field redefinition for the SME}

\author{Ralf Lehnert}

\email[Electronic mail: ]{rlehnert@lns.mit.edu}
\affiliation{Center for Theoretical Physics,
Massachusetts Institute of Technology, Cambridge, MA 02139}

\date{September 22, 2006}

\begin{abstract}
This work instigates a study of non-local field mappings 
within the Lorentz- and CPT-violating Standard-Model Extension (SME). 
An example of such a mapping is constructed explicitly, 
and the conditions for the existence of its inverse are investigated. 
It is demonstrated 
that the associated field redefinition 
can remove $b^{\mu}$-type Lorentz violation from free SME fermions 
in certain situations. 
These results are employed 
to obtain explicit expressions for the corresponding Lorentz-breaking momentum-space eigenspinors 
and their orthogonality relations.
\end{abstract}

\pacs{11.30.Cp, 11.30.Er, 12.60.-i}

\maketitle

\section{Introduction}
\label{introduction}

Despite its numerous phenomenological successes, 
the framework of the Standard Model coupled to general relativity 
is not believed to provide a fundamental description of nature; 
it is rather viewed as the low-energy limit 
of some encompassing quantum theory 
in which the gravitational, strong, and electroweak interactions are unified. 
The energy scale of such an encompassing theory 
is expected to be of the order of
the Planck mass $M_{Pl}\simeq10^{19}\,$GeV. 
This expectation presents an enormous experimental challenge 
because the emerging effects 
are likely to be suppressed 
by one or more powers of $M_{Pl}$ 
at presently attainable energies. 
However, 
minute Lorentz and CPT breakdown 
has recently been identified as a promising signal in this context: 
such effects may arise in various approaches to fundamental physics, 
and they are amenable to ultrahigh-precision tests \cite{cpt04,reviews}. 

The low-energy effects of Lorentz and CPT breakdown 
are described by an effective-field-theory 
framework called the Standard-Model Extension (SME) \cite{ck97,sme,gravext}. 
Besides the usual Standard-Model and Einstein--Hilbert actions, 
this framework contains all leading-order contributions 
to the action 
that violate particle Lorentz and CPT symmetry 
while maintaining coordinate independence. 
Of particular phenomenological interest 
is the minimal SME. 
In addition to conventional physics, 
it only contains those Lorentz- and CPT-violating terms 
that are expected to be dominant 
and satisfy a few other physically desirable requirements. 
The Minkowski-spacetime limit of the minimal SME 
has been the focus of various investigations, 
including ones 
with photons \cite{photonexpt,radiative,photonth2,km02}, 
electrons \cite{trap,spinpol,eexpt3}, 
protons and neutrons \cite{cc,spaceexpt,bnsyn}, 
mesons \cite{hadronexpt}, 
muons \cite{muexpt}, 
neutrinos \cite{neutrinos,nuexpt}, 
and the Higgs \cite{higgs}. 
Bounds in the gravity sector have also been obtained recently \cite{bk06,shore}. 

The Lorentz- and CPT-violating coefficients in the minimal SME 
are non-dynamical background vectors and tensors, 
which are coupled to Standard-Model fields. 
They are assumed to be generated by underlying physics. 
Various specific mechanisms 
for such effects 
have been proposed in the literature. 
For instance, 
mechanisms compatible with spontaneous Lorentz and CPT breakdown 
have been studied in models based on 
string theory 
\cite{ksp,gyb}, 
noncommutative geometry 
\cite{ncqed}, 
spacetime-varying fields 
\cite{spacetimevarying,ghostcond}, 
quantum gravity 
\cite{qg}, 
nontrivial spacetime topology 
\cite{klink}, 
random-dynamics models 
\cite{fn02}, 
multiverses 
\cite{bj}, 
and brane-world scenarios 
\cite{brane}. 

There are a number of known mappings between SME fields 
that relate different SME coefficients for Lorentz violation \cite{ck97,sme,km02,redef}. 
In some cases, 
the effects of Lorentz breakdown 
can be moved from one SME sector to another. 
In other cases, 
the coefficient can be removed from the SME altogether 
implying it is unobservable. 
Moreover, 
for certain dimensionless Lorentz-violating parameters for fermions 
a spinor redefinition is needed to obtain a hermitian Hamiltonian. 
It therefore follows 
that such field redefinitions play an important role 
for the analysis of experimental Lorentz and CPT tests 
and for the interpretation of Lorentz violation in the SME. 
The form of possible field redefinitions 
is essentially only constrained by 
the requirement of invertibility. 
On one hand, 
this leaves a substantial amount of freedom in 
the identification of useful field redefinitions. 
On the other hand, 
the large number of possibilities 
hampers a systematic and comprehensive study of such redefinitions. 

Previously analyzed mappings between SME fields 
fall into two categories \cite{ck97,sme,km02,redef}: 
redefinitions of the field variables 
and coordinate rescalings. 
In both cases, 
the mapping is local 
and applies on- as well as off-shell. 
The present work is intended to launch an investigation 
of a set of field redefinitions 
characterized by non-locality. 
We will focus on 
a mapping within the SME's free $b^{\mu}$ model  
that scales 
(and hence eliminates in a certain limit) 
the $b^{\mu}$ coefficient from this model. 
With the exception of special cases, 
the structure of the non-local mapping is such 
that only on-shell applicability is guaranteed. 

Although the $b^{\mu}$ coefficient 
cannot be removed from a realistic interacting model, 
this field redefinition is nevertheless interesting 
for the following reasons. 
In many situations, 
the extraction of the physical content of a field theory, 
such as the SME, 
requires an initial investigation of its quadratic sectors. 
For instance, 
perturbation theory in quantum field theory 
typically amounts to an expansion about known free-field physics. 
Since our non-local field redefinition 
can be employed to establish a one-to-one correspondence 
between the usual free Dirac field 
and the non-interacting $b^{\mu}$ model, 
it can be used to gain insight into the free-field physics 
of this model. 
In particular, 
this idea allows a complete 
characterization of the solutions to the free $b^{\mu}$ model. 
The present work illustrates this with a specific example: 
the previously unknown $b^{\mu}$ eigenspinors 
are generated from the conventional ones 
via our non-local field redefinition. 
We finally remark 
that a generalization of this idea to the full SME, 
if possible,  
would completely characterize its free sectors.

The paper is organized as follows. 
A brief review of the $b^{\mu}$ model 
is given in Sec.\ \ref{model}. 
Section \ref{plvector} discusses the Pauli--Lubanski vector, 
which is an essential ingredient in our non-local field redefinition. 
In Secs.\ \ref{shiftoperator} and \ref{Rinverse}, 
the field redefinition is constructed 
and some of its properties including invertibility are established. 
Section \ref{eigespinors} 
employs these results 
for the determination 
of explicit expressions for the $b^{\mu}$-model 
momentum-space eigenspinors. 
A summary and outlook 
is contained in Sec.\ \ref{sum}. 
Throughout we employ the notation and conventions 
of Ref.\ \cite{kl01}.

\section{Review: fermions with  a ${\bm \gamma_5 /\hspace{-1.6mm}b}$ term}
\label{model}

This section 
reviews various results derived in Ref.\ \cite{ck97} 
on the relativistic quantum mechanics 
of spin-$\half$ fermions 
with a $\gamma_5 /\hspace{-1.7mm}b$ term. 
The starting point is the Lagrangian 
\beq{blagrangian}
{\cal L}_b=
\overline{\psi}\left(\half i/\hspace{-2.75mm}
\stackrel{\leftrightarrow}{\hspace{-1mm}\partial}\hspace{-1.75mm}{}
-m-\gamma_5 /\hspace{-1.7mm}b\right)\psi\;,
\eeq
where $m$ is the fermion's mass. 
The last term of this Lagrangian 
contains a nondynamical spacetime-constant vector $b^\mu$.
This vector coefficient has mass dimensions,
and it explicitly breaks Lorentz and CPT symmetry.
The ordinary Dirac Lagrangian ${\cal L}_0$
is recovered for zero $b^\mu$.

Lagrangian \rf{blagrangian} has recently 
received substantial attention 
in Lorentz-violation research.
It has been studied 
in the context of radiative corrections in Lorentz-breaking electrodynamics \cite{radiative},
finite-temperature field theory \cite{finiteT}, 
as well as particle-decay processes \cite{decay},
and it might be generated within the ghost-condensate model \cite{ghostcond}.
We also mention 
that experiments with spin-polarized matter 
have constrained the order of magnitude of $b^\mu$ 
to be less than $10^{-30}\,$GeV for electrons \cite{spinpol}. 
Additional bounds of $10^{-27}\,$GeV for both electrons and protons, 
as well as $10^{-31}\,$GeV for neutrons  
have been obtained by clock-comparison tests \cite{cc}. 
The clean limit of $10^{-25}\,$GeV 
on a component of $b^\mu$ for the electron
has been extracted from Penning-trap experiments \cite{trap}. 
Under certain assumptions, 
the order of magnitude of $b^\mu$ 
is less than $10^{-20}\,$GeV 
for neutrinos \cite{neutrinos}. 
Throughout this work, 
we shall thus take $b^2\ll m^2$. 

The modified Dirac equation 
determined by Lagrangian \rf{blagrangian} 
is 
\beq{bequation}
\left(i/\hspace{-1.87mm}
\partial\hspace{-.2mm}{}
-m-\gamma_5 /\hspace{-1.7mm}b\right)\psi_b(t,\vec{r})=0\;,
\eeq
which can be rearranged 
to give the Schr\"odinger equation 
$i\partial_0\psi_b(t,\vec{r})=H_b\,\psi_b(t,\vec{r})$ 
with the hermitian Hamiltonian 
\beq{bhamiltonian} 
H_b=\gamma^0(-i\;\vec{\gamma}\sdot\vec{\nabla}+m+\gamma_5 /\hspace{-1.7mm}b)\;.
\eeq
We denote the 
space 
of solutions $\psi_b(t,\vec{r})$ 
to Eq.\ \rf{bequation} 
by ${\cal S}_b$.
In the $b^\mu=0$ limit, 
the ordinary Dirac case with solutions $\psi_0(t,\vec{r})$ 
and solution space ${\cal S}_0$ emerges. 
For later convenience, 
we abbreviate the modified Dirac operator 
appearing above by 
\beq{DiracOpDef} 
D_b\equiv i/\hspace{-1.87mm}\partial\hspace{-.2mm}{}
-m-\gamma_5 /\hspace{-1.7mm}b\;.
\eeq
Reversing the sign of the mass term in $D_b$
and applying the resulting operator 
from the left 
to Eq.\ \rf{bequation} 
yields the following modified Klein--Gordon equation:
\beq{bkgequation}
\left(\Box+m^2+b^2+2\gamma_5 \sigma^{\mu\nu}b_\nu\partial_{\nu}\right)\psi_b=0\;.
\eeq
With a second operator-squaring procedure, 
one can derive the following equation 
diagonal in spinor space: 
\beq{4equation}
\left[(\Box+m^2+b^2)^2-4b^2\Box+4(b\sdot\partial)^2\right]\psi_b=0\;.
\eeq
A plane-wave ansatz $w(\lambda)\exp(-ix\sdot\lambda)$ for $\psi_b(x)$
in the above equation yields
\beq{bdr}
(\lambda^2-m^2-b^2)^2+4b^2\lambda^2-4(\lambda\sdot b)^2=0
\eeq
for the fermion's dispersion relation. 
For any fixed wave vector $\vec{\lambda}$, 
this dispersion relation constitutes a quartic equation 
in the plane-wave frequency. 
We denote its 
four roots by 
$(\lambda^{\pm}_a)^{\mu}=\big((\lambda^{\pm}_a)^{0}(\vec{\lambda}),\vec{\lambda}\big)$. 
These roots are associated with fermion (superscript $+$) and antifermion (superscript $-$) solutions, 
each with two possible spin states ($a=1,2$). 
In Sec.\ \ref{plvector}, 
we will see 
that $a$ labels spins parallel and antiparallel to $b^\mu$. 
Explicit expressions for $(\lambda^{\pm}_a)^{0}(\vec{\lambda})$ 
are given in Appendix \ref{roots}. 
Jointly with Eq.\ \rf{bequation}, 
the dispersion relation \rf{bdr} 
determines the corresponding eigenspinors $w^{\pm}_a(\vec{\lambda})$. 
Most analyses in this work 
do not require 
a full distinction between all the roots. 
For notational convenience, 
we then only display the necessary labels.

\section{The Pauli--Lubanski vector} 
\label{plvector} 

Up to the momentum-operator factor $\sqrt{P^{\mu}P_{\mu}}$, 
the Pauli--Lubanski vector $W^{\mu}$ 
is the relativistic generalization 
of a particle's spin. 
For Dirac fermions, 
this vector is given in position space by \cite{IZ} 
\beq{plv} 
W^\mu = \half\gamma_5\sigma^{\mu\nu}\partial_\nu\;. 
\eeq 
Although Lorentz breakdown 
precludes the conservation of total angular momentum 
in the $b^\mu$ model \rf{blagrangian}, 
$W^{\mu}W_{\mu}=\half(\half+1)\Box$ 
is evidently conserved:
its gamma-matrix structure is trivial, 
and so it commutes with $H_b$. 
We remark in passing 
that this argument does not employ the modified Dirac Eq.\ \rf{bequation}. 
Hence, 
$[H_b,W^2]=0$ is not only valid on solutions $\psi_b(t,\vec{r})$ 
but actually on any sufficiently well-behaved spinor $\psi(t,\vec{r})$. 
This essentially means 
that the Lorentz-violating $b^\mu$ interaction 
leaves unchanged the spin-$\half$ character 
of the particle $\psi_b$, 
as expected. 
Note 
that the above argument only uses 
the spacetime independence of the Hamiltonian. 
The result therefore generalizes to fermions
with translation-invariant Lorentz violation 
in the full SME, 
which incorporates the minimal SME.

In the usual Dirac case, 
an analogous argument 
establishes the conservation of the individual spin components $\sim\!W^{\mu}$. 
In the present case,  
such an argument fails 
due to the absence of Lorentz symmetry. 
However, 
Lorentz transformations in the plane orthogonal to $b^\mu$ 
still determine a symmetry of the Lagrangian \rf{blagrangian}, 
leading us to investigate $\Wb$ 
as a candidate for a conserved spin component. 

We begin by analyzing the eigenvalues of $\Wb$. 
Let $\chi \exp(-ix\sdot\lambda)$ be an arbitrary momentum eigenspinor  
that does not necessarily obey an equation of motion, 
i.e., 
both $\chi$ and $\lambda^{\mu}$ are unconstrained. 
Any eigenvalue $\Omega$ 
then satisfies $\det(-\half i \gamma_5\sigma^{\mu\nu}b_{\mu}\lambda_{\nu}-\Omega)=0$. 
We can obtain an explicit expression 
for the square of this determinant 
as follows. 
Note that $\det(M)=\det(CMC^{-1})$ 
for any $M$ and any invertible $C$, 
so that $\det(M)^2=\det(MCMC^{-1})$. 
In our case, $M=-\half i \gamma_5\sigma^{\mu\nu}b_{\mu}\lambda_{\nu}-\Omega$. 
If now $C$ is chosen to be the usual charge-conjugation matrix 
(e.g., $C=i\gamma^2\gamma^0$ in the Dirac representation), 
a diagonal expression for $MCMC^{-1}$ emerges. 
This yields a positive and a negative eigenvalue, 
both twofold degenerate: 
\beq{genev} 
\Omega_{a}=\half(-1)^{a}\sqrt{(\lambda\sdot b)^2-b^2\lambda^2}\;, 
\eeq 
where $a=1,2$.
It follows 
that $\Wb$ and $P^{\mu}$ have simultaneous eigenstates $\chi_a(\lambda)\exp(-ix\sdot\lambda)$. 

In order to employ the general result \rf{genev} 
within the $b^\mu$ model \rf{blagrangian} and the ordinary Dirac case, 
we must still verify 
that $\Wb$ commutes with $H_b$ and $H_0$, 
respectively. 
As a first step, 
note that $[D_b,\Wb]=0$. 
We remark in passing 
that this shows $\Wb\;{\cal S}_b\subset {\cal S}_b$, 
a necessary condition for our claim. 
Employing $D_b=\gamma^0(i\partial_0-H_b)$ 
in this commutator gives after some algebra 
\beq{commutator} 
[H_b,\Wb]=\gamma^0[\gamma^0,\Wb]D_b\;.
\eeq 
It is apparent 
that this commutator 
is in general nonzero, 
when it acts on arbitrary spinors $\psi(t,\vec{r})$. 
However, 
we only need $[H_b,\Wb]\,\psi_b(t,\vec{r})=0$ 
for all $\psi_b(t,\vec{r})$ satisfying Eq.\ \rf{bequation}, 
which is ensured by the presence of the Dirac operator $D_b$ 
on the right-hand side of Eq.\ \rf{commutator}. 
Thus, 
on the solution space ${\cal S}_b$ we indeed have 
\beq{rcommutator} 
[H_b,\Wb]_{{\cal S}_b}=0\;. 
\eeq 
A similar result holds for $H_0$. 
We can therefore conclude 
that $\Wb$ is conserved. 
In particular, 
simultaneous energy--momentum eigenspinors  of $W \!\cdot b$ and $H_b$
as well as simultaneous energy--momentum eigenspinors of $W \!\cdot b$ and $H_0$ 
exist. 
The eigenvalue formula \rf{genev} 
is thus applicable in the $b^\mu$ model \rf{blagrangian} 
and in the ordinary Dirac case.

We may now employ 
the appropriate dispersion relations 
to reduce the general eigenvalue expression \rf{genev} 
in each of the two specific cases: 
for the $b^\mu$ model, 
the dispersion relation \rf{bdr} yields 
\beq{bev} 
\Omega_a=\pm\frac{1}{4}(-1)^a(\lambda_a^2-m^2-b^2)\;,
\eeq 
and for the usual Dirac field, 
$\lambda^2_a=m^2$ gives  
\beq{diracev} 
\Omega_a=\half(-1)^a\sqrt{(\lambda_a\sdot b)^2-m^2b^2}\;.
\eeq 
Note 
that in the $b^{\mu}$ case 
the correspondence of the signs is left open. 
This issue can be resolved as follows. 
The last term in the modified Klein--Gordon equation \rf{bkgequation} 
is equal to $4\,\Wb$. 
With this observation, 
the momentum-space version of Eq.\ \rf{bkgequation} becomes 
\beq{kgmeq} 
(\lambda^2-m^2-b^2-4\Wb)w(\lambda)=0\;. 
\eeq 
This fixes the sign ambiguity in Eq.\ \rf{bev}, 
and one can now write 
\beq{bevfixed} 
\Omega_a=\frac{1}{4}(\lambda_a^2-m^2-b^2) 
\eeq 
for the eigenvalues of $\Wb$ 
in the $b^\mu$ model. 
We remark 
that this argument also provides 
an independent proof of the fact 
that the momentum eigenspinors of $H_b$ 
are at the same time 
eigenstates of $\Wb$. 

The gamma-matrix structure of $\Wb$ 
is determined by $\sigma^{\mu\nu}$. 
Only the $\sigma^{jk}$, 
where Latin indices range from 1 to 3, 
are hermitian, 
so the question arises 
as to whether $W\!\cdot b$ is observable, 
at least in principle. 
It can be answered 
with the help of the eigenvalues \rf{diracev} and \rf{bevfixed}. 
For the $b^{\mu}$ model, 
the eigenvalues \rf{bevfixed} of $W\!\cdot b$ are real 
because the hermiticity of $H_b$ implies $\lambda^2\in\mathbb{R}$. 
This is compatible 
with the observability of $\Wb$ in models with nonzero $b^{\mu}$.
For the ordinary Dirac field, 
only the second term under the square root in Eq.\ \rf{diracev} 
together with a timelike $b^\mu$
can potentially lead to complex $\Omega_a$. 
However, 
in a coordinate system in which $b^{\mu}=(B,\vec{0})$ 
one can verify 
that $\Omega_a^2\ge0$. 
It follows 
that also in the conventional Dirac model 
the eigenvalues of $\Wb$ are consistent 
with the observability of this operator. 

Another question concerns the inverse of $\Wb$. 
There are situations in which $\Omega_a=0$ and no inverse exists, 
for example, 
if $\lambda^{\mu}$ is parallel to $b^\mu$. 
This issue is analyzed in more detail 
in Appendix \ref{invert}. 
But if we exclude such special cases, 
we can determine $(\Wb)^{-1}$. 
We begin  
by observing 
that $(\sigma_{\mu\nu}A^{\mu}B^{\nu})^2=A^2B^2-(A\sdot B)^2$ 
for any two 4-vectors $A^{\mu}$ and $B^{\nu}$. 
Thus, 
in momentum space we obtain 
\bea{msinvwb} 
(\Wb)^{-1}&=&\frac{-2i\,\gamma_5\sigma^{\mu\nu}b_{\mu}\lambda_{\nu}}{(\lambda\sdot b)^2-\lambda^2b^2}\nonumber\\
&=&\frac{4\,\Wb}{(\lambda\sdot b)^2-\lambda^2b^2}\;.
\eea 
In position space, 
we formally write 
\beq{psinvwb} 
(\Wb)^{-1}=\frac{4\,\Wb}{b^2\Box-(b\sdot\partial)^2}\;, 
\eeq 
where the action of the inverse derivative-type operator 
on any (well-behaved) position-space function in $f$
is defined explicitly by 
\beq{invdo1} 
\frac{1}{b^2\Box-(b\sdot\partial)^2}\;f(x)\equiv
\int\! d^4y\;G(x-y)\;f(y)\;. 
\eeq
Here, the Green function $G$ is given by 
\beq{green} 
G(x)=\int_{\cal C}\frac{d^4\lambda}{(2\pi)^4}\frac{e^{-i\lambda\cdot x}}{(\lambda\sdot b)^2-\lambda^2b^2}\;,
\eeq 
as usual. 
If $(\Wb)^{-1}$ acts on function spaces on which it can become singular 
(i.e., $\Wb=0$), 
the freedom in the choice of the contour ${\cal C}$ 
may be used to select certain boundary conditions. 
In the present work, 
we will only need to consider 
the action of $(\Wb)^{-1}$ on ${\cal S}_0$ or ${\cal S}_b$ for $b^2\leq 0$. 
The discussion in Appendix \ref{invert} demonstrates 
that $(\Wb)^{-1}$ is nonsingular 
in these situations. 
Then, 
the contour simply runs along the real-$\lambda^0$ axis, 
and any ambiguities in the selection of ${\cal C}$ are absent. 

Finally, 
consider $\int\!d^4y\;G(x-y)\;\partial f(y)/\partial y^{\mu}$ 
and integrate by parts. 
If $f$ falls off sufficiently fast 
and is sufficiently smooth,
the boundary term can be dropped, 
we may trade $\partial/\partial y^{\mu}$ for $-\partial/\partial x^{\mu}$, 
and then pull this derivative outside the integral. 
This shows explicitly 
that $\Wb$ commutes with $[b^2\Box-(b\sdot\partial)^2]^{-1}$, 
as expected for two spacetime-independent expressions 
containing only the momentum operator.  
It follows 
that there are no ordering ambiguities in Eq.\ \rf{psinvwb}.

\section{The ${\bm b^{\bm \mu}}$-shift operator ${\cal{\bm  R}}_{\bm \xi}$}
\label{shiftoperator}

Our goal is 
to find an explicit representation of an operator ${\cal R}_\xi(x)$ 
that maps solutions of the $b^\mu$ model with coefficient $b^\mu$ 
to solutions of another $b^\mu$ model with shifted coefficient $b^{\mu}+\xi b^{\mu}$, 
where $\xi\in\mathbb{R}$.
In other words, 
we seek an operator 
\beq{RProp} 
{\cal R}_\xi: {\cal S}_b\rightarrow{\cal S}_{b+\xi b}\;, 
\eeq 
where the size (but not the direction) of $b^\mu$ is changed. 
Only when ${\cal R}_\xi$ acts on solutions of the usual Dirac equation $b^{\mu}=0$, 
the above stipulation is meaningless 
and must be amended. 
In this special case, 
it is necessary 
to first select a $b^\mu$ vector 
that is nonzero but can otherwise be arbitrary. 
We then require ${\cal R}_\xi$ to generate solutions of a model with coefficient $\xi b_{\mu}$, 
so that ${\cal R}_\xi: {\cal S}_0\rightarrow{\cal S}_{\xi b}$. 
For notational simplicity, 
we have suppressed the vector character of various subscripts. 

{\bf Definition.} 
An operator with the properties of ${\cal R}_\xi$ indeed exists: 
\beq{rdef} 
{\cal R}_\xi(x)=\exp\left(-\xi x_{\mu}\vec{B}^{\mu\nu}\vec{\partial}_{\nu}\right)\;, 
\eeq 
where the projector-type quantity $\vec{B}^{\mu\nu}$ 
is given by 
\beq{Bdef} 
\vec{B}^{\mu\nu}=\frac{b^2\eta^{\mu\nu}-b^{\mu}b^{\nu}}{2\,\Wb}\;. 
\eeq 
Here, 
the coefficient $b^{\mu}\neq 0$ 
is that of the space ${\cal S}_{b}$ 
the operator ${\cal R}_\xi$ acts on. 
The only exception, 
noted in the previous paragraph, 
is the special case ${\cal R}_\xi\,{\cal S}_0$, 
in which $b^\mu$ can be chosen freely. 
At this stage, 
the mathematical meaning of Eqs.\ \rf{rdef} and \rf{Bdef} 
is somewhat vague. 
In the remainder of this section, 
we make the above definition more precise 
and establish key properties of ${\cal R}_\xi$. 

The exponential in Eq.\ \rf{rdef} 
is to be understood as a short-hand notation 
for the power-series expansion 
\beq{Rseries} 
\exp\left(-\xi x_{\mu}\vec{B}^{\mu\nu}\vec{\partial}_{\nu}\right)\equiv
\sum_{n=0}^\infty\frac{1}{n!}\left(-\xi x_{\mu}\vec{B}^{\mu\nu}\vec{\partial}_{\nu}\right)^n\;. 
\eeq 
The derivative $\vec{\partial}_{\nu}$ 
in this expression 
is to be taken 
with respect to $x_{\mu}$, 
the position-space variable. 
Since the commutator $[\vec{\partial}_{\nu},x_{\mu}]=\eta_{\mu\nu}$ is nonzero 
and products of $x$ and $\partial$ appear above, 
their order must be specified. 
A similar issue arises for $\vec{B}^{\mu\nu}$ and $x_{\mu}$ 
because $\vec{B}^{\mu\nu}$ contains 
the position-space expression for $\Wb$. 
In Eq.\ \rf{Rseries}, 
the operator ordering is defined such 
that none of the derivatives and integrations 
are acting on the $x_{\mu}$s. 
The $n$th term in the series looks therefore as follows: 
\beq{nRseries} 
\frac{\xi^n}{n!}\left(x_{\mu}\vec{B}^{\mu\nu}\vec{\partial}_{\nu}\right)^n\equiv
\frac{\xi^n}{n!}\prod_{j=1}^n x_{\mu_j}\prod_{k=1}^n \vec{B}^{\mu_j\nu_k}\vec{\partial}_{\nu_k}\;.
\eeq
Note 
that the operator order 
in the second product on the right-hand side of Eq.\ \rf{nRseries} 
is irrelevant: 
$\vec{B}^{\mu\nu}$ and derivatives commute 
since $\Wb$ and $\partial$ have simultaneous eigenstates. 
As a reminder for the ordering \rf{nRseries}, 
$\vec{\partial}_{\nu}$ and $\vec{B}^{\mu\nu}$ carry arrows 
indicating the direction of action. 

The definition of $\vec{B}^{\mu\nu}$ 
contains $(\Wb)^{-1}$, 
which may be singular in certain cases. 
Since $\vec{B}^{\mu\nu}$ appears always contracted with a 4-gradient, 
we may address this issue 
by considering $\vec{B}^{\mu\nu}\vec{\partial}_{\nu}$ instead. 
Also, 
our interest lies in mappings between models 
with parallel $b^\mu$ coefficients of different length, 
including the case $b^{\mu}=0$. 
It therefore suffices 
to specify the action of the above combined operator 
on a complete set spanning 
${\cal S}_b$ and ${\cal S}_0$. 
We select the set of plane-wave eigenspinors 
$\psi^{\lambda}_a(x)=w_a(\vec{\lambda})\exp(-ix\sdot\lambda_a)$ of $\Wb$, 
where $a=1,2$ labels the eigenvalues of $\Wb$, 
as before. 
This gives 
\beq{Baction} 
\vec{B}^{\mu\nu}\partial_{\nu}\psi^{\lambda}_a(x) 
=-iB^{\mu\nu}_a(\vec{\lambda})\;(\lambda_a)_{\nu}\;\psi^{\lambda}_a(x)\;, 
\eeq 
where we have introduced the momentum-space version of $\vec{B}^{\mu\nu}$:
\beq{Bdef} 
B^{\mu\nu}_a(\vec{\lambda})\equiv
(-1)^{a}\frac{b^2\eta^{\mu\nu}-b^{\mu}b^{\nu}}{\sqrt{(\lambda_a\sdot b)^2-b^2\lambda_a^2}}\;.
\eeq 
To arrive at this result, 
we have used Eq.\ \rf{genev}. 

Note 
that $B^{\mu\nu}_a(\vec{\lambda})$ becomes singular, 
when the square root vanishes. 
In Appendix \ref{invert}, 
we show 
that this requires
$\lambda^{\mu}$ to be parallel to a timelike $b^\mu$: 
$(\lambda^\pm_a)^{\mu}(\vec{\lambda}_0)=\zeta b^{\mu}$, 
where $\zeta$ is a dimensionless constant, 
$b^2>0$, 
and $\vec{\lambda}_0$ is the location of the singularity. 
The most natural and straightforward definition of $B^{\mu\nu}_a(\vec{\lambda}_0)$ 
would employ the limit $\vec{\lambda}\to\vec{\lambda}_0$ at the $\Wb$ singularity:
\beq{singularitydef} 
B^{\mu\nu}_a(\vec{\lambda}_0) \;\lambda_{\nu}
\equiv\lim_{\vec{\lambda}\to\vec{\lambda}_0} 
B^{\mu\nu}_a(\vec{\lambda})\;\lambda_{\nu}\;. 
\eeq 
The remaining task is to determine the limit explicitly. 
This is simplified 
by working a coordinate system 
in which $b^{\mu}=(B,\vec{0})$. 
We can decompose any 4-momentum as 
$p^{\mu}=\lambda^{\mu}(\vec{\lambda}_0)+\varsigma b^{\mu}+\varepsilon u^{\mu}$, 
where $\varsigma$ and $\varepsilon$ are parameters 
and $u^{\mu}=(0,\vec{u})$ obeys $u^2=-1$. 
Employing this expression 
in definition \rf{singularitydef} 
where we have to take $\varsigma,\varepsilon\to0$ yields 
\beq{singularity} 
B^{\mu\nu}_a(\vec{\lambda}_0) \;\lambda_{\nu}
=(-1)^a|B|\;u^{\mu}\;. 
\eeq 
Although this result is finite, 
the presence of $u^{\mu}$ indicates 
that the limit depends upon the path
by which $(\lambda^\pm_a)^{\mu}(\vec{\lambda}_0)$ is approached. 
This non-uniqueness means 
that an inverse of $\Wb$ 
is ambiguous 
for those states characterized 
at the beginning of this paragraph. 
Some results in the subsequent sections 
require $\Wb$ 
to be invertible, 
so that they are only valid 
when these states are excluded.  
Note, 
however,
that this issue only arises for timelike $b^\mu$ 
and only for a subset of measure zero 
in the respective ${\cal S}_b$. 

{\bf Useful properties.} 
We next establish two basic properties of ${\cal R}_{\xi}(x)$. 
The first of these properties concerns the action 
of ${\cal R}_{\xi}(x)$ on plane-wave spinors $\psi^{\lambda}_a(x)$ 
introduced earlier. 
Starting from the power-series definition \rf{Rseries}, 
it is apparent 
that the gradients $\vec{\partial}_{\mu}$ 
(including those in the denominator of $\vec{B}^{\mu\nu}$) 
can be replaced by $-i\lambda_{\mu}$ 
when acting on $\psi^{\lambda}_a(x)$. 
The resulting expression 
contains no longer derivatives, 
operator ordering becomes irrelevant, 
and the series can be summed: 
\beq{Raction} 
{\cal R}_{\xi}(x)\;\psi^{\lambda}_a(x)
=\exp\big(i\xi x_{\mu}B_a^{\mu\nu}(\lambda_a)_{\nu}\big)\;\psi^{\lambda}_a(x)\;.
\eeq

The second property 
concerns the derivative of ${\cal R}_{\xi}(x)$. 
Beginning again with the series \rf{Rseries}, 
one can verify that 
\beq{Rderivative} 
\left[\partial^{\mu}{\cal R}_{\xi}(x)\right]=
-{\cal R}_{\xi}(x)\xi\vec{B}^{\mu\nu}\vec{\partial}_{\nu}\;.
\eeq 
This essentially means 
that the symbolic ``exp'' in Eq.\ \rf{rdef} 
behaves as a true exponential 
with regards to differentiation. 
Note, 
however, 
that the operator ordering matters. 
One can also show  
that 
\beq{prule} 
\partial^{\mu}({\cal R}_{\xi}\psi) 
=(\partial^{\mu}{\cal R}_{\xi})\psi 
+{\cal R}_{\xi}(\partial^{\mu}\psi)\;, 
\eeq 
i.e., 
the usual product rule applies, 
as expected. 

{\bf Proof of Relation \rf{RProp}.} 
We are now in the position 
to establish the initial claim 
that ${\cal R}_\xi$ changes 
the magnitude of $b^\mu$. 
It has to be verified 
that 
\beq{proposition} 
D_{b+\xi b}\,{\cal R}_\xi\,\psi_b=0\quad 
\textrm{if} \quad D_{b}\,\psi_b=0\;, 
\eeq 
where $D_{b}$ and $D_{b+\xi b}$ are Dirac operators 
defined by Eq.\ \rf{DiracOpDef}. 
We first use 
that $W\!\cdot b$ 
(which determines the gamma-matrix structure of ${\cal R}_\xi$) 
and $D_{b+\xi b}$ 
commute. 
Displaying the spinor indices $c$, $d$, and $f$ for clarity, 
one obtains 
$(D_{b+\xi b})_{cd}({\cal R}_\xi)_{df}=
(D_{b+\xi b})_{df}({\cal R}_\xi)_{cd}$. 
Since $D_{b+\xi b}$ contains the derivative 
$i/\hspace{-1.87mm}\partial\hspace{-.2mm}{}$, 
the product rule \rf{prule} generates 
an additional term 
when $(D_{b+\xi b})_{df}$ is moved past $({\cal R}_\xi)_{cd}$. 
With the result \rf{Rderivative} at hand, 
we then obtain 
\beq{step2}
(D_{b+\xi b})_{df}({\cal R}_\xi)_{cd}=
({\cal R}_\xi)_{cd}(D_{b+\xi b}
-i\xi\gamma_{\mu}\vec{B}^{\mu\nu}\vec{\partial}_{\nu})_{df}\;. 
\eeq
The position of the spinor indices 
is now such 
that we may convert back to matrix notation. 
Moreover, 
it can be verified 
that $i\gamma_{\mu}\vec{B}^{\mu\nu}\vec{\partial}_{\nu}=-\xi\gamma_5 /\hspace{-1.7mm}b$ 
on any sufficiently well-behaved spinor $\psi(t,\vec{r})$. 
This finally yields 
\beq{step3}
D_{b+\xi b}\,{\cal R}_\xi\,\psi_b={\cal R}_\xi(D_{b+\xi b}+\xi\gamma_5 /\hspace{-1.7mm}b)\,\psi_b=D_{b}\,\psi_b=0\;, 
\eeq
where we have employed the modified Dirac equation $D_{b}\,\psi_b=0$ in the last step. 
This demonstrates that 
\beq{result} 
{\cal R}_\xi\,\psi_b=\psi_{b+\xi b}\;, 
\eeq
i.e., 
the operator ${\cal R}_\xi$ 
maps any solution of a model with Lorentz-violating coefficient $b^\mu$ 
to some solution of a model with coefficient $b^{\mu}+\xi b^{\mu}$.

\section{Inverse of the ${\bm b^{\bm \mu}}$-shift operator}
\label{Rinverse}

Thus far, 
we have found 
that ${\cal R}_\xi\,{\cal S}_b\subset{\cal S}_{b+\xi b}$. 
The goal of this section is 
to sharpen this statement. 
We will establish 
that ${\cal R}_\xi$ determines, 
in fact, 
a one-to-one correspondence 
between the elements of ${\cal S}_b$ 
and those of ${\cal S}_{b+\xi b}$. 
Then, 
the inverse of ${\cal R}_\xi$ exists, 
and a number of useful insights and applications can be established.
For example, 
certain properties and relations derived within a model with a specific $b^\mu$ coefficient 
can be mapped to analogous results for other models with more general $b^\mu$. 

The basic idea behind establishing the bijectivity of ${\cal R}_\xi$ 
is the following. 
Both the range ${\cal S}_b$ and the domain ${\cal S}_{b+\xi b}$ 
are spanned by the plane-wave eigenspinors of the respective Dirac equations. 
If we can show 
that for each eigenspinor in ${\cal S}_b$ 
there is exactly one eigenspinor in ${\cal S}_{b+\xi b}$, 
${\cal R}_\xi$ is one-to-one. 
If we can further demonstrate 
that each ${\cal S}_{b+\xi b}$ eigenspinor 
can be obtained via this mapping, 
${\cal R}_\xi$ is onto, 
and the claim follows. 
We will establish this result in three steps. 
In the first step, 
we show that eigenspinors are, 
in fact, 
mapped to eigenspinors. 
The second step verifies 
that ${\cal R}_\xi$ does not mix the four branches of eigenspinors. 
This roughly means 
that particles (antiparticles) are mapped to particles (antiparticles) 
such that their spin state is left unaffected. 
As the final third step, 
we demonstrate 
that for each of the resulting four maps between branches 
the plane-wave momentum is mapped one-to-one and onto. 

{\bf Eigenspinors are mapped to eigenspinors.} 
Let 
$\psi^{\lambda}_a(x)=w_a^{b}(\vec{\lambda})\exp(-ix\cdot\lambda_a)$. 
Here, 
the spinor superscript $b$ 
refers to the $b^\mu$ case. 
With Eq.\ \rf{Raction}, 
one can establish 
that ${\cal R}_\xi$ 
inserts an additional plane-wave exponential into the expression for $\psi^{\lambda}_a$: 
\beq{Rbasis} 
{\cal R}_\xi\;\psi^{\lambda}_a
=w_a^{b}(\vec{\lambda})\exp\big(i\xi x_{\mu} B^{\mu\nu}_a (\lambda_a)_{\nu}\big)\exp(-ix\sdot\lambda_a)\;. 
\eeq 
Combining the exponentials shows 
that  
\beq{ldef} 
(\Lambda_{a'})^{\mu}\equiv(\lambda_a)^{\mu}-\xi B^{\mu\nu}_a(\lambda_a)_{\nu} 
\eeq 
must be interpreted as the new plane-wave momentum. 
Since $(\lambda_a)_{\mu}$ satisfies the dispersion relation \rf{bdr}, 
$(\Lambda_{a'})^{\mu}$ is constrained as well: 
one can verify 
that it also satisfies Eq.\ \rf{bdr}, 
but with $b^\mu$ replaced by $b^{\mu}+\xi b^{\mu}$, 
as expected. 

We next use the fact \rf{result} 
that ${\cal R}_\xi\psi^{\lambda}_a\in{\cal S}_{b+\xi b}$, 
which gives
\beq{rel1}
[i/\hspace{-1.87mm}\partial\hspace{-.2mm}{}
-m-(1+\xi)\gamma_5 /\hspace{-1.7mm}b]\;
w_a^{b}(\vec{\lambda})\exp(-ix\sdot\Lambda_{a'})=0\;.
\eeq
It therefore follows 
that in addition to its defining relation 
$[/\hspace{-2.12mm}\lambda_a\hspace{-.2mm}{}
-m-\gamma_5 /\hspace{-1.7mm}b]\;
w_a^{b}(\vec{\lambda})=0$, 
the momentum-space spinor $w_a^{b}(\vec{\lambda})$ 
also obeys 
$[/\hspace{-2.1mm}\Lambda_{a'}\hspace{-.2mm}{}
-m-(1+\xi)\gamma_5 /\hspace{-1.7mm}b]\;
w_a^{b}(\vec{\lambda})=0$. 
But this is the definition of $w^{b+\xi b}_{a'}(\vec{\Lambda})$. 
We therefore have 
\beq{Rspinor} 
w_{a'}^{b+\xi b}(\Lambda)={\cal R}_{\xi}\;w_a^{b}(\lambda)\;, 
\eeq  
where $\Lambda_{a'}$ and $\lambda_{a}$ 
are related by Eq.\ \rf{ldef}. 
The above results 
lead to the  conclusion 
that ${\cal R}_{\xi}$ 
maps plane-wave eigenspinors of $\Wb$ for a model with coefficient $b^\mu$ 
into those for a model with coefficient $b^{\mu}+\xi b^{\mu}$. 
The proof 
that the map ${\cal R}_\xi:{\cal S}_b\rightarrow{\cal S}_{b+\xi b}$ is a bijection
is now reduced to the following. 
We have to show 
that each plane-wave eigenspinor in ${\cal S}_b$ 
corresponds to exactly one plane-wave eigenspinor in ${\cal S}_{b+\xi b}$ 
and that this correspondence is onto. 

{\bf Branches are mapped to branches.} 
Any solution space ${\cal S}_b$ 
contains four distinct branches of eigenspinors 
labeled 
by the sign of the plane-wave frequency $(\lambda^\pm_a)^0(\vec{\lambda})$
at large wave vectors 
\footnote{For $b^{\mu}\neq 0$, 
all dispersion-relation branches 
have regions with spacelike $(\lambda^\pm_a)^{\mu}$, 
and one can always select frames 
in which part of a positive-frequency branch 
dips below the frequency zero 
and vice versa \cite{kl01}. 
However, 
the sign of the {\em asymptotic} plane-wave frequencies 
$(\lambda^\pm_a)^0(|\vec{\lambda}|\to\infty)$ 
remains unchanged 
under any finite coordinate boost.} 
and the sign of the $\Wb$ eigenvalue $\Omega_a$. 
In other words, 
there are the usual particle and antiparticle solutions, 
each with two possible spin projections along $b^{\mu}$, 
as discussed in Sec.\ \ref{model}. 
In what follows, 
we will show 
that ${\cal R}_\xi$ maps branches to branches 
without mixing them. 
More precisely, 
the image of an ${\cal S}_b$ branch 
lies on one and only one ${\cal S}_{b+\xi b}$ branch; 
the images of any two distinct ${\cal S}_b$ branches 
belong to distinct ${\cal S}_{b+\xi b}$ branches. 
To this end, 
we need to investigate 
the behavior of the sign of $(\lambda^\pm_a)^0(\vec{\lambda})$ 
and the sign of $\Omega_a$ 
under the mapping ${\cal R}_\xi$. 

We first consider the sign of the plane-wave frequency. 
Its behavior under ${\cal R}_\xi$ 
is determined by Eq.\ \rf{ldef}. 
For timelike $b^{\mu}$, 
we immediately find difficulties. 
The projector-type quantity $B^{\mu\nu}_a$ appearing in Eq.\ \rf{ldef} 
contains $(\Wb)^{-1}$, 
which may not exist for certain $\vec{\lambda}$, 
as discussed in Sec.\ \ref{shiftoperator}. 
For lightlike $b^\mu$, 
on the other hand, 
such issues are absent. 
Equation \rf{ldef} reduces to 
\beq{lighlike_map} 
(\Lambda^{\pm}_{a'})^{\mu}=
(\lambda^{\pm}_a)^{\mu}-(-1)^a{\rm sgn}(b\cdot\!\lambda^{\pm}_a)\,\xi\, b^{\mu}\; .
\eeq 
Up to a sign, 
${\cal R}_\xi$ just adds the constant vector $b^{\mu}$ to $(\lambda^{\pm}_a)^{\mu}$. 
For large $\vec{\lambda}$,  
we thus have ${\rm sgn}(\Lambda^{\pm}_{a'})^{0}={\rm sgn}(\lambda^{\pm}_a)^{0}$, 
which justifies the $\pm$ label on $(\Lambda^{\pm}_{a'})^{\mu}$. 

For spacelike $b^\mu$, 
we may select coordinates such that $b^{\mu}=(0,\vec{B})$. 
Then, Eq.\ \rf{ldef} becomes 
\beq{spacelike_map} 
(\Lambda^{\pm}_{a'})^{\mu}=
(\lambda^{\pm}_a)^{\mu}+(-1)^a\xi\fr{\vec{B}^2(\lambda^{\pm}_a)^{\mu}-(\vec{\lambda}\sdot\vec{B})\,b^{\mu}}
{\sqrt{(\vec{\lambda}\sdot\vec{B})^2+\vec{B}^2{\lambda^\pm_a}^2}}\; .
\eeq
The second term on the right-hand side of Eq.\ \rf{spacelike_map} 
has the structure $(\lambda^\pm_a)_{\nu}\,\epsilon^{\mu\nu}$, 
where we have defined the tensor 
\beq{etensor} 
\epsilon_{\mu\nu}\equiv(-1)^a\xi\frac{\vec{B}^2\eta^{\mu\nu}+b^{\mu}b^{\nu}}
{\sqrt{(\vec{\lambda}\cdot\vec{B})^2+\vec{B}^2{\lambda^\pm_a}^2}}\;. 
\eeq
The $\pm$ label on $(\Lambda^{\pm}_{a'})^{\mu}$ is justified, 
if the components of $\epsilon^{\mu\nu}$ are small compared to 1. 
The dispersion relation \rf{spacelike_roots} yields 
\beq{drresult}
{\lambda^\pm_a}^2=m^2+B^2+2(-1)^a\sqrt{m^2B^2+(\vec{\lambda}\sdot\vec{B})^2}\;. 
\eeq
Then, 
the minimum of the square root in Eq.\ \rf{etensor} is given by $|\vec{B}|\big(m+(-1)^a|\vec{B}|\big)$, 
where we have used our assumption $|b^2|\ll m^2$. 
It follows 
that the components of $\epsilon_{\mu\nu}$ 
are ${\cal O}\big(|\vec{B}|/m\big)\ll 1$, 
which establishes the desired result. 

We have seen above 
that the ${\cal R}_\xi$ map leaves unchanged the $\pm$ label of the plane-wave frequencies, 
at least for lightlike and spacelike $b^{\mu}$. 
We now need to study the behavior of the $a$ label 
under this map. 
The eigenvalue equation for $\Omega_a$ 
reads 
\beq{eqfev1} 
\left[\frac{i}{2} \gamma_5\sigma_{\mu\nu}b^{\mu}(\lambda^{\pm}_{a})^{\nu}+\Omega_a\right]w^{\pm}_a(\vec{\lambda})=0\;.
\eeq 
Since ${\cal R}_\xi$ just inserts an additional plane-wave exponential 
into the position-space eigenspinors,  
the eigenvalue equation changes under ${\cal R}_\xi$ 
according to $(\lambda^{\pm}_{a})^{\mu}\to(\Lambda^{\pm}_{a'})^{\mu}$. 
In the case 
when ${\cal R}_\xi$ connects two models with nontrivial $b^{\mu}$, 
we also need to take $b^{\mu}\to(1+\xi)\,b^{\mu}$ 
in the expression for $\Wb$. 
Then, the mapped eigenvalue $\Omega'_{a'}$ is given by 
\beq{eqfev2} 
\left[\frac{i}{2} \gamma_5\sigma_{\mu\nu}(1+\xi)b^{\mu}(\Lambda^{\pm}_{a'})^{\nu}
+\Omega'_{a'}\right]w^{\pm}_a(\vec{\lambda})=0\;. 
\eeq
Comparison of the two eigenvalue equations \rf{eqfev1} and \rf{eqfev2} implies 
\beq{a_map} 
\Omega'_{a'}=(1+\xi)\left(\Omega_{a}-\half\,\xi\, b^2\right)\;, 
\eeq 
where we have used Eq.\ \rf{ldef}. 
The value of $(1+\xi)$ maybe positive or negative, 
but it is fixed and does not change, e.g., along on a branch. 
So if we can show 
that $\Omega_a$ dominates the right-hand side of Eq.\ \rf{a_map}, 
it will determine the sign of $\Omega'_{a'}$. 
In the lightlike $b^{\mu}$ case, 
this feature is clear, 
and for timelike $b^{\mu}$, 
the aforementioned difficulties arise. 
The discussion in the previous paragraph implies 
$\Omega_a\ge\half|\vec{B}|\big(m-|\vec{B}|\big)$ 
for $b^{\mu}=(0,\vec{B})$. 
It follows 
that the dominance of $\Omega_a$ is also 
assured for spacelike $b^{\mu}$. 

The above results establish 
that ${\cal R}_\xi$ maps an $a$ branch  
to a single other branch. 
But one might wonder 
why the other branch has a different label $a'\neq a$ 
when $(1+\xi)<0$. 
The explanation for this fact simple. 
With respect to a fixed $b^{\mu}$, 
the spin alignment actually remains fixed under ${\cal R}_\xi$, 
i.e., the spin stays either parallel or antiparallel to $b^{\mu}$. 
However, 
${\cal R}_\xi$ replaces $b^{\mu}\to(1+\xi)\,b^{\mu}$, 
so that the projection axis 
(and not the spin) 
reverses direction 
for $(1+\xi)<0$. 
This leads to different signs 
for $\Omega'_{a'}$ and $\Omega_{a}$. 

As mentioned above, 
we must slightly modify Eq.\ \rf{a_map}
when either the domain or the range of the ${\cal R}_\xi$ map 
involves the space ${\cal S}_0$ of solutions to the conventional Dirac equation. 
For example, 
$\xi =-1$ maps a model with a nontrivial $b^\mu$ 
to the usual Dirac case, 
but then the right-hand side of Eq.\ \rf{a_map} vanishes. 
This issue arises 
due the mapping $\Wb\to(1+\xi)\,\Wb$ 
in the above derivation of Eq.\ \rf{a_map}. 
For mappings between ${\cal S}_0$ and ${\cal S}_b$, 
we may instead chose to leave $\Wb$ unchanged. 
We then obtain 
\beq{a_map_special} 
\Omega'_{a'}=\Omega_{a}\pm\half\, b^2\;, 
\eeq 
where the upper and lower signs 
refer to the cases with ${\cal S}_0$ 
as the range or domain, 
respectively. 

{\bf Momenta map is bijective for each branch.}
As claimed, 
we have demonstrated above 
that ${\cal R}_\xi$ maps branches to branches, 
at least for lightlike and spacelike $b^{\mu}$. 
This essentially 
establishes the invertibility of ${\cal R}_\xi$ 
with regards to the spinor degrees of freedom. 
It remains to study the $\vec{\lambda}$-momentum degrees of freedom,  
a task 
that can now be performed branch by branch. 
The momentum map is given by Eq.\ \rf{ldef}, 
and we need to show 
that it is onto and invertible. 
The timelike $b^{\mu}$ case must again be excluded. 
For a lightlike $b^{\mu}$, 
Eq.\ \rf{lighlike_map} emerges and clearly shows 
that the map is onto. 
The map also implies 
that $b\cdot\!\lambda^{\pm}_a=b\cdot\!\Lambda^{\pm}_{a'}$, 
which ensures invertibility. 

For spacelike $b^{\mu}$, 
we need to study the second term on the right-hand side of Eq.\ \rf{spacelike_map}, 
which is given by $\epsilon^{\mu\nu}(\lambda^\pm_a)_{\nu}$. 
We have established earlier 
that $\epsilon^{\mu\nu}$ is a correction suppressed by at least $|\vec{B}|/m$. 
Moreover, 
$\epsilon^{\mu\nu}$ is smooth, 
so that the map \rf{spacelike_map} must be onto. 
When $(\lambda^{\pm}_{a})^{\mu}$ satisfies the usual dispersion relation ${\lambda^\pm_a}^2=m^2$, 
the Jacobian for the map \rf{spacelike_map} 
is given by 
\beq{jacobian} 
\left|\frac{\partial (\Lambda_a^\pm)^i}{\partial\lambda^j}\right|
=\left(1+(-1)^a\frac{\vec{B}^2}{\sqrt{(\vec{\lambda}\sdot\vec{B})^2+m^2\vec{B}^2}}\right)^2. 
\eeq 
Since this Jacobian is strictly nonzero, 
the map \rf{spacelike_map} is invertible, 
and thus bijective in this situation. 
More general mappings $b^{\mu}\to(1+\xi)\,b^{\mu}$ for nonzero $b^{\mu}$ 
can always be decomposed as $b^{\mu}\to 0 \to (1+\xi)\,b^{\mu}$, 
where each step is bijective by the above result. 
It follows 
that bijectivity is also guaranteed 
for arbitrary spacelike $b^{\mu}$. 

{\bf Explicit expression for inverse map.}
The above analysis has shown 
that for both lightlike and spacelike $b^{\mu}$ 
the mapping generated by ${\cal R}_\xi$ is bijective. 
This implies ${\cal R}_\xi$ has an inverse $R^{-1}_\xi$. 
In cases not involving the usual Dirac model with ${\cal S}_0$, 
it is natural to expect 
that $R^{-1}_\xi=R_{\xi'}$, 
where $\xi'$ is defined by $(1+\xi')(1+\xi)=1$: 
\beq{inverse} 
({\cal R}_\xi)^{-1}=R_{-\xi/(1+\xi)}\;. 
\eeq 
In situations with ${\cal S}_0$, 
either as the domain or the range, 
we anticipate 
\bea{inverse_special} 
(R_1)^{-1}&=&R_{-1}\;,\nonumber\\
(R_{-1})^{-1}&=&R_{1}\;, 
\eea 
where $R_1$ 
generates ${\cal S}_b$ from the conventional ${\cal S}_0$.
We may verify Eqs.\ \rf{inverse} and \rf{inverse_special} 
by demonstrating 
that $(R)^{-1}R\;\psi^{\lambda}_{a\pm}=\psi^{\lambda}_{a\pm}$ 
holds for any plane-wave eigenspinor 
$\psi^{\lambda}_{a\pm}(x)=w_a^{\pm}(\vec{\lambda})\exp(-ix\sdot\lambda^{\pm}_a)$ 
in the appropriate ${\cal S}_b$ or ${\cal S}_0$. 
In other words, 
we have to show 
that the frequency label $\pm$, 
the spin label $a$, 
and the wave 4-vector $\lambda^\pm_a$ 
remain unchanged under $(R)^{-1}R$. 

We first note 
that the frequency label $\pm$ 
is indeed unaffected by the map ${\cal R}_\xi$
for any $\xi$, 
as established earlier in this section. 
If either the domain or the range of ${\cal R}_\xi$ 
involves the conventional Dirac space ${\cal S}_0$, 
the result \rf{a_map_special} shows 
that the spin label $a$ 
is also left unaffected by ${\cal R}_\xi$. 
In all other cases, 
Eq.\ \rf{a_map} holds. 
The twofold iteration of this equation 
appropriate in the present situation 
proves 
that the label $a$ is invariant under $R_{-\xi/(1+\xi)}\,{\cal R}_\xi$, 
as desired. 

For the plane-wave vector, 
we want to invert
\beq{ltransform_special} 
(\Lambda_{a})^{\mu}=(\lambda_a)^{\mu}\pm
\frac{b^2(\lambda_a)^{\mu}-(\lambda_a\sdot\, b)\,b^{\mu}}{2\Omega_a}\;, 
\eeq 
in situations involving the usual Dirac case. 
Here, the upper (lower) sign 
refers to the case with ${\cal S}_0$ 
as the range (domain). 
One can check that indeed
\beq{linvtransform_special} 
(\lambda_{a})^{\mu}=(\Lambda_a)^{\mu}\mp
\frac{b^2(\Lambda_a)^{\mu}-(\Lambda_a\sdot\, b)\,b^{\mu}}{2\Omega'_a}\;, 
\eeq 
as anticipated. 
For all other maps, 
i.e., those not involving the conventional Dirac model, 
we seek to invert  
\beq{ltransform} 
(\Lambda_{a})^{\mu}=(\lambda_a)^{\mu}-
\xi\,\frac{b^2(\lambda_a)^{\mu}-(\lambda_a\sdot\, b)\,b^{\mu}}{2\Omega_a}\;. 
\eeq 
One can again verify 
that our expectation 
\beq{linvtransform} 
(\lambda_{a})^{\mu}=(\Lambda_a)^{\mu}+\frac{\xi}{1+\xi}\,
\frac{b^2(\Lambda_a)^{\mu}-(\Lambda_a\sdot\, b)\,b^{\mu}}{2\Omega'_{a'}} 
\eeq 
is correct, 
which establishes Eqs.\ \rf{inverse} and \rf{inverse_special}.

{\bf Bottom line.}
For lightlike and spacelike $b^\mu$ 
we have now explicit expressions 
for a $b^{\mu}$-shift operator $R_{\xi}$ and its inverse. 
This operator connects solutions 
of two $b^{\mu}$ models 
with parameters $b^\mu$ and $(1+\xi)\,b^\mu$ 
in a one-to-one fashion. 
In particular, 
it is possible to map a $b^{\mu}$ model 
to the conventional Dirac case, 
and vice versa. 
A composition of two $b^{\mu}$-shift operators 
can thus be used 
to map a $b^{\mu}$ model 
to a $\tilde{b}^{\mu}$ model 
via $b^{\mu}\to 0 \to \tilde{b}^\mu$ 
for {\em any} coefficients $b^2,\tilde{b}^2\leq 0$. 
We remark 
that even though a single ${\cal R}_\xi$ maps 
a plane-wave eigenspinor in one model 
to a single plane-wave eigenspinor in another model, 
the above composition of ${\cal R}_\xi$ operators 
will typically generate a linear superposition of eigenspinors. 
This arises because a single ${\cal R}_\xi$ only scales 
the spin-quantization axis $b^{\mu}$, 
whereas a composition involving ${\cal S}_0$ 
can change the direction of $b^{\mu}$, 
and thus lead to new spin-projection states. 

The ${\cal R}_\xi$ map can be viewed as a field redefinition. 
An example of another field redefinition discussed in the literature \cite{ck97} 
is $\psi(x)\to e^{-ia\cdot x}\psi(x)$. 
This field redefinition does not only remove $a^{\mu}$ from the free $a^{\mu}$ model, 
but also from one-flavor QED. 
Such a generalization is possible 
because $\psi(x)\to e^{-ia\cdot x}\psi(x)$ has two key properties. 
First, 
the $a^{\mu}$ field redefinition is also defined off-shell. 
Second, 
it leaves the current $j^{\mu}=\overline{\psi}\gamma^{\mu}\psi$, 
and thus the coupling to electrodynamics, 
unchanged. 
Our ${\cal R}_\xi$ field redefinition 
does not seem to exhibit these properties in general. 
First, 
the action of ${\cal R}_\xi$ is only defined on ${\cal S}_b$, 
where $b^{\mu}$ is lightlike, spacelike, or vanishing. 
But interactions would require an off-shell extension of the ${\cal R}_\xi$ map. 
However, 
such an extension may face obstacles 
similar to those encountered 
in Secs.\ \ref{shiftoperator} and \ref{Rinverse} for timelike $b^{\mu}$. 
Second, 
the current $j^{\mu}=\overline{\psi}\gamma^{\mu}\psi$, 
and thus the coupling to electrodynamics, 
is altered by ${\cal R}_\xi$. 
It therefore follows 
that $b^{\mu}$ cannot be removed from QED. 
It is, 
in fact, 
physical and can in principle be measured. 
However, 
${\cal R}_\xi$ does have other applications in the free $b^{\mu}$ model, 
one of which is discussed in the next section.

\section{Explicit eigenspinors} 
\label{eigespinors}

The $b^\mu$-model generalizations of various key features of the usual free Dirac case are known. 
Instances of these are the dispersion relation, 
the energy--momentum tensor, 
certain expressions for spinor projectors, 
and the Feynman propagator \cite{ck97,rl04}. 
The determination of other generalizations 
is often hampered by the complexity 
that arises through the inclusion of Lorentz violation. 
The momentum eigenspinors for the $b^\mu$ model 
are one such example. 
Besides approximations, 
only the eigenspinors for $b^\mu=(B,\vec{0})$ 
have been obtained \cite{ck97}. 
In this section, 
we determine the momentum eigenspinors 
for lightlike and spacelike $b^\mu$. 
To this end, 
we employ the ${\cal R}_\xi$ operator 
to map the known eigenspinors in the conventional Dirac case 
to those of the desired $b^\mu$ model. 

{\bf Conventional eigenspinors.} 
The conventional momentum-space eigenspinors 
obey $(/\hspace{-2.1mm}\lambda^\pm_a - m)w^\pm_a (\vec{\lambda})=0$. 
We may take $w^\pm_a(\vec{\lambda})=(/\hspace{-2.1mm}\lambda^\pm_a + m)W^\pm_a$, 
where $W^\pm_a$ is an arbitrary spinor \cite{IZ}. 
Since we have 
$(/\hspace{-2.1mm}\lambda^\pm_a - m)(/\hspace{-2.1mm}\lambda^\pm_a + m)={\lambda^\pm_a}^2-m^2=0$, 
this ansatz satisfies the above defining equation for $w^\pm_a(\vec{\lambda})$. 
The choice 
\beq{Wspinor} 
W^+_a=\left(\begin{array}{c} 
\phi_a\\0\\0
\end{array}\right)\;,
\quad W^-_a=\left(\begin{array}{c} 
0\\0\\ \chi_a
\end{array}\right) 
\eeq
ensures that $w^\pm_a(\vec{\lambda})$ is nonzero. 
Here, 
$\phi_a$ and $\chi_a$ are non-vanishing, 
but otherwise arbitrary two-component spinors. 
In the Dirac representation for the gamma matrices, 
we obtain explicitly 
\bea{conventional_spinor} 
w^+_a(\vec{\lambda})&=&\frac{1}{\sqrt{2m(E+m)}}
\left(\begin{array}{c} 
(E+m)\,\phi_a\\ (\vec{\lambda}\sdot\vec{\sigma})\,\phi_a
\end{array}\right)\;,\nonumber\\
w^-_a(\vec{\lambda})&=&\frac{1}{\sqrt{2m(E+m)}}
\left(\begin{array}{c} 
-(\vec{\lambda}\sdot\vec{\sigma})\,\chi_a\\ (E+m)\,\chi_a
\end{array}\right)\;, 
\eea
where we have set $E=(\lambda^+_a)^0=-(\lambda^-_a)^0=\sqrt{m^2+\vec{\lambda}^2}$. 
The factor $1/\sqrt{2m(E+m)}$ 
has been included for normalization. 
These spinors satisfy the orthogonality relations 
\beq{orthonorm1} 
{w^\pm_a}^\dagger(\vec{\lambda})\, w^\mp_{a'}(\vec{\lambda})=0\;,
\quad{w^\pm_a}^\dagger(\vec{\lambda})\, w^\pm_{a'}(\vec{\lambda})=\frac{E}{m}\,\delta_{aa'}\;, 
\eeq 
if $\phi_a$ and $\chi_a$ are chosen 
such that ${\phi_{a}}^{\!\!\dagger}\phi_{a'}={\chi_a}^{\!\!\dagger}\chi_{a'}=\delta_{aa'}$. 
The analogous relations 
involving the Dirac conjugate spinors $\overline{w}^\pm_a = {w^\pm_a}^\dagger\gamma^0$ 
are 
\beq{orthonorm2} 
\overline{w}^\pm_a(\vec{\lambda})\, w^\mp_{a'}(-\vec{\lambda})=0\;,
\quad\overline{w}^\pm_a(\vec{\lambda})\, w^\pm_{a'}(-\vec{\lambda})=\pm\delta_{aa'}\;. 
\eeq 
We remark in passing 
that the second of these equations 
may also be written as  
$\overline{w}^\pm_a(\vec{\lambda})\,w^\pm_{a'}(\vec{\lambda})=\pm\delta_{aa'}$. 
Our sign choice in Eq.\ \rf{orthonorm2} 
becomes the natural one 
after the usual reinterpretation of the negative-energy solutions. 

We intend to map the conventional spinors \rf{conventional_spinor} 
to a Lorentz-violating model with coefficient $b^\mu=(b^0,\vec{b})$. 
It is therefore convenient 
to use the remaining freedom in $\phi_a$ and $\chi_a$
to require the $w^\pm_{a}(\vec{\lambda})$ spinors 
to be eigenstates of 
$\Wb$. 
For the positive-frequency solutions, 
the upper two components of the $\Wb$ eigenvalue equation give 
\beq{pos_freq_Wb} 
(\vec{n}_+\!\cdot\vec{\sigma})\,\phi_a=(-1)^a\,\phi_a\;.
\eeq
Here, we have used Eq.\ \rf{diracev}, and we have defined 
\beq{lplus} 
\vec{n}_+ \equiv \frac{m(E+m)\,\vec{b}-(\lambda\sdot b+b^0m)\,\vec{\lambda}}
{\sqrt{(\lambda\sdot b)^2-m^2b^2}\,(E+m)}\;.
\eeq 
One can check 
that $\vec{n}_+$ has unit length 
and that the  equation for the lower two components 
can be derived from Eq.\ \rf{pos_freq_Wb}, 
as required by consistency. 
It follows 
that $\phi_a$ must be the eigenvector $\phi_a(\vec{n}_+)$ 
of $\vec{n}_+\!\cdot\vec{\sigma}$ 
with eigenvalue $(-1)^a$. 
If the spherical-polar angles 
that $\vec{n}_+$ subtends are specified as $(\theta,\varphi)$, 
then we have explicitly: 
\beq{explicitphi} 
{}\hspace{-1.5mm}\phi_1(\vec{n}_+)=\left(\begin{array}{c} 
e^{-i\varphi}\sin \frac{\theta}{2}\\ -\cos \frac{\theta}{2}
\end{array}\right),
\hspace{1.5mm}\phi_2(\vec{n}_+)=\left(\begin{array}{c} 
\cos \frac{\theta}{2}\\e^{i\varphi}\sin \frac{\theta}{2}
\end{array}\right).
\eeq 
An analogous reasoning for the negative-frequency spinors 
yields $\chi_a=\chi_a(\vec{n}_-)$, 
where 
\beq{nminus} 
\vec{n}_-\equiv -\frac{m(E+m)\,\vec{b}-(\lambda\sdot b-b^0m)\,\vec{\lambda}}
{\sqrt{(\lambda\sdot b)^2-m^2b^2}\,(E+m)}\;.
\eeq 

{\bf Eigenspinors for ${\bm b^{\bm \mu}}$ model.}
We can now employ ${\cal R}_\xi$ 
to map these conventional momentum-space eigenspinors 
to those for a $b^\mu$ model. 
Equation \rf{Rspinor} shows 
that the conventional spinors also satisfy the $b^\mu$ model, 
but at a different momentum 
determined by ${\cal R}_\xi$. 
Hence, 
the remaining task is to express the conventional-case $\vec{\lambda}$ 
in terms of the $b^\mu$-model momentum $\Lambda^{\pm}_{a}$. 
The appropriate transformations are the lower-sign relations 
in Eqs.\ \rf{ltransform_special} and \rf{linvtransform_special}. 
To determine compact and explicit expressions for the ${\cal S}_b$ momentum-space spinors, 
we write Eq.\ \rf{linvtransform_special} in the form
\beq{compact} 
\lambda^{\pm}_{a}=\Lambda^{\pm}_{a}+\delta^{\pm}_{a}\;,
\eeq
where
\beq{deltadef}
(\delta^{\pm}_{a})^{\mu} \equiv (-1)^a\,
\frac{b^2(\Lambda^{\pm}_a)^{\mu}-(\Lambda^{\pm}_a\sdot\, b)\,b^{\mu}}{\sqrt{(\Lambda^{\pm}_a\sdot\, b)^2-b^2{\Lambda^{\pm\,2}_a} }}\;.
\eeq
The $b^\mu$-model momentum-space spinors $W^\pm_a(\vec{\Lambda})$ 
are then given by $W^\pm_a(\vec{\Lambda})=w^\pm_a(\vec{\Lambda}+\vec{\delta}^{\pm}_{a})$. 
We explicitly obtain for these spinors:
\begin{widetext}
\beq{bmu_spinors} 
W^+_a(\vec{\Lambda})=
\frac{1}{\sqrt{2m}}\!\left(\renewcommand{\arraystretch}{2.5}
\begin{array}{c} 
\sqrt{m+(\Lambda^{+}_{a})^{0}+(\delta^{+}_{a})^{0}}\;\phi_a(\vec{N}^a_+)\\ 
\frac{\displaystyle (\vec{\Lambda}+\vec{\delta}^{+}_{a})\sdot\vec{\sigma}}
{\displaystyle \sqrt{m+(\Lambda^{+}_{a})^{0}+(\delta^{+}_{a})^{0}}}\;\phi_a(\vec{N}^a_+)
\end{array}\right),\quad
W^-_a(\vec{\Lambda})=\frac{1}{\sqrt{2m}}\!
\left(\renewcommand{\arraystretch}{2.5}
\begin{array}{c} 
\frac{\displaystyle -(\vec{\Lambda}+\vec{\delta}^{-}_{a})\sdot\vec{\sigma}}
{\displaystyle \sqrt{m-(\Lambda^{-}_{a})^{0}-(\delta^{-}_{a})^{0}}}\;\chi_a(\vec{N}^a_-)\\ 
\sqrt{m-(\Lambda^{-}_{a})^{0}-(\delta^{-}_{a})^{0}}\;\chi_a(\vec{N}^a_-)
\end{array}\right). 
\eeq 
Here, 
the vectors $\vec{N}^a_+$ and $\vec{N}^a_-$ are given by 
\beq{Nvector} 
\vec{N}^a_\pm= \pm\frac{m\big[m\pm (\Lambda^{\pm}_{a})^{0}\pm(\delta^{-}_{a})^{0}\big]\vec{b}
-(\Lambda^{\pm}_a\sdot b\pm b^0m)(\vec{\lambda}+\vec{\delta}^{\pm}_{a})}
{\sqrt{(\Lambda^{\pm}_a\sdot b)^2-m^2b^2}\,\big(m\pm (\Lambda^{\pm}_{a})^{0}\pm(\delta^{-}_{a})^{0}\big)}\;.
\eeq 
\end{widetext} 
For lightlike $b^{\mu}$, 
we have $(\delta^{\pm}_a)^{\mu}=-(-1)^a\,{\rm sgn}(\Lambda^{\pm}_a\cdot b)\,b^{\mu}$, 
so that in this case the $b^{\mu}$-model momentum-space spinors 
take a relatively simply form. 

{\bf Orthogonality relations in the ${\bm b^{\bm \mu}}$ model.}
Let us finally comment on 
the orthogonality relations for $W^\pm_a(\vec{\Lambda})$. 
Employing Eq.\ \rf{ltransform_special}, 
we may express $\vec{\Lambda}$ in terms of $\vec{\lambda}$ and write 
\beq{Weq}
W^\pm_a(\vec{\lambda}-\vec{\kappa}^\pm_a)=w^\pm_a(\vec{\lambda})\;, 
\eeq
where 
\beq{kappadef} 
(\kappa^{\pm}_{a})^{\mu}\equiv(-1)^a\,
\frac{b^2(\lambda^{\pm}_a)^{\mu}-(\lambda^{\pm}_a\sdot\, b)\,b^{\mu}}{\sqrt{(\lambda^{\pm}_a\sdot\, b)^2-m^2 b^2 }}\;.
\eeq 
For notational consistency, 
the we rename $\vec{\lambda}\to\vec{\Lambda}$, 
which entails $(\lambda^{\pm}_a)^0\to\pm\sqrt{m^2+\vec{\Lambda}^2}\neq(\Lambda^{\pm}_a)^0$.
In what follows, 
the dependence of $\kappa^{\pm}_{a}$ 
on $\vec{\Lambda}$ is understood. 
The results of the mapping 
of the conventional orthogonality relations \rf{orthonorm1} and \rf{orthonorm2} 
to the $b^{\mu}$-model relations 
are now given by: 
\bea{orthonorm1mapped} 
\!\!\!\!\!\!\!\!{W^\pm_a}^\dagger(\vec{\Lambda}-\vec{\kappa}^\pm_a)\, 
W^\mp_{a'}(\vec{\Lambda}-\vec{\kappa}^\mp_{a'})&=&0\;,\nonumber\\ 
\!\!\!\!\!\!\!\!{W^\pm_a}^\dagger(\vec{\Lambda}-\vec{\kappa}^\pm_a)\, 
W^\pm_{a'}(\vec{\Lambda}-\vec{\kappa}^\pm_{a'})&=&\frac{\sqrt{m^2+\vec{\Lambda}^2}}{m}\,\delta_{aa'}\;, 
\eea 
and 
\bea{orthonorm2mapped} 
\!\!\!\!\!\!\overline{W}^\pm_a(\vec{\Lambda}-\vec{\kappa}^\pm_{a,\vec{\Lambda}})\, 
W^\mp_{a'}(-\vec{\Lambda}-\vec{\kappa}^\mp_{a',-\vec{\Lambda}})&=&0\;,\nonumber\\
\!\!\!\!\!\!\overline{W}^\pm_a(\vec{\Lambda}-\vec{\kappa}^\pm_{a,\vec{\Lambda}})\, 
W^\pm_{a'}(-\vec{\Lambda}-\vec{\kappa}^\pm_{a',-\vec{\Lambda}})&=&\pm\delta_{aa'}\;. 
\eea 
For clarity, 
we have made explicit the dependence of $\kappa^{\pm}_{a}$ 
on $\vec{\Lambda}$ in Eq.\ \rf{orthonorm2mapped}. 

The four non-vanishing relations in the second line of Eq.\ \rf{orthonorm1mapped} 
involve scalar products of two spinors with the {\em same} momentum argument. 
It follows 
that in these four equations we may shift the momentum arguments 
to obtain simpler, more conventional expressions. 
However, 
the other orthogonality relations 
involve spinors with {\em differing} momentum arguments. 
The question arises 
if these relations would also hold 
at equal momentum arguments. 
For Eq.\ \rf{orthonorm2mapped}, 
this is not the case \cite{ck97}. 
On the other hand, 
the vanishing relations in Eq.\ \rf{orthonorm1mapped} 
do possess an equal-argument analogue: 
they are eigenvectors of the momentum-space Hamiltonian, 
and as such they are orthogonal for any fixed $\vec{\Lambda}$. 
A possible degeneracy of the eigenenergies 
does not invalidate this conclusion. 
With our assumption $b^2\ll m^2$, 
such a degeneracy is impossible 
for eigenenergies with differing $\pm$ labels. 
For energy degeneracies between states with differing $a$ label, 
the corresponding spinors are orthogonal 
by virtue of being eigenvectors of $\Wb$. 
We thus have 
\bea{orthonorm1mapped2} 
\!\!\!\!\!\!\!\!{W^\pm_a}^\dagger(\vec{\Lambda})\, 
W^\mp_{a'}(\vec{\Lambda})&=&0\;,\nonumber\\ 
\!\!\!\!\!\!\!\!{W^\pm_a}^\dagger(\vec{\Lambda})\, 
W^\pm_{a'}(\vec{\Lambda})&=&\frac{\sqrt{m^2+(\vec{\Lambda}+\vec{\kappa}^\pm_a)^2}}{m}\,\delta_{aa'}\;. 
\eea 
We remark 
that the above induced spinor normalization 
differs from the choice in Refs.\ \cite{ck97,kl01,rl04}. 
This is, however, 
acceptable 
because observables do not depend on 
the choice of normalization \cite{ck01}. 
We also note 
that starting from Eq.\ \rf{orthonorm1mapped2} and mapping back to ${\cal S}_0$ 
produces orthogonality relations 
for the conventional Dirac case. 
They are nontrivial, 
i.e., involve differing momentum arguments, 
for the vanishing relations in Eq.\ \rf{orthonorm1mapped2}.

\section{Summary and outlook} 
\label{sum} 

The present work 
has initiated the study of a novel type of field redefinitions 
within the Lorentz- and CPT-violating SME. 
As opposed to previously known SME field redefinitions, 
the mappings considered here involved infinitely many derivatives 
and integrations, 
a feature associated with non-locality. 

In the context of the Lorentz-violating $b^{\mu}$ model, 
we have constructed a non-local operator ${\cal R}_\xi$, 
given by Eq.\ \rf{rdef}, 
that induces an on-shell field redefinition, 
such that the redefined field 
satisfies an equation of motion with a scaled coefficient $(1+\xi)\,b^{\mu}$. 
For lightlike and spacelike $b^{\mu}$, 
this field redefinition is bijective. 
The special choice $\xi=-1$ 
therefore permits the removal of Lorentz violation 
from the free $b^{\mu}$ model for $b^2\leq 0$. 
As for other field redefinitions considered in the literature, 
${\cal R}_\xi$ cannot in general be applied in interacting models.   

The significance of the ${\cal R}_\xi$ map 
lies in the fact, 
that it establishes a one-to-one correspondence 
between the Lorentz-violating $b^{\mu}$ model 
and the conventional Dirac case for $b^2\leq 0$. 
This permits the determination and complete characterization 
of the solutions to the $b^{\mu}$ model, 
which is a prerequisite for many phenomenological studies 
including perturbation theory. 
As an example of this idea, 
we have constructed the previously unknown explicit 
momentum-space eigenspinors of the $b^{\mu}$ model 
and their generalized orthogonality relations. 
These results are contained in 
Eq.\ \rf{bmu_spinors} and Eqs.\ \rf{orthonorm1mapped}--\rf{orthonorm1mapped2}. 

The present work has opened several avenues for further research. 
For instance, 
it would be interesting to establish 
whether ${\cal R}_\xi$ 
possesses an off-shell extension. 
Such an extension would allow theoretical studies in an interacting model. 
Another example for future investigations 
is the determination of ${\cal R}_\xi$-type on-shell maps 
for other Lorentz-violating SME coefficients. 
This would yield 
the full characterization of solutions 
to the corresponding free SME sector. 
Finally, 
the spinor analysis in Sec.\ \ref{eigespinors} 
could be extended to other conventional features 
and their Lorentz-violating analogues in the $b^{\mu}$ model.

\acknowledgments
The author thanks Roman Jackiw for discussions. 
This work is supported 
by the U.S.\ Department of Energy 
under cooperative research agreement No.\ DE-FG02-05ER41360 
and by the European Commission 
under Grant No.\ MOIF-CT-2005-008687.

\appendix

\section{Dispersion-relation roots} 
\label{roots} 

The dispersion relation \rf{bdr} 
constitutes a fourth-order equation 
in the plane-wave frequency $\lambda^{0}$. 
In the three canonical coordinate systems 
associated with a lightlike, spacelike, and timelike $b^{\mu}$, 
the roots of this equation 
take a relatively simple form \cite{ck97}. 
Moreover, 
the particle (antiparticle) interpretations of these roots 
are known \cite{ck97}. 
Here, 
we list these results 
and adjust the $a$ label 
to be consistent with our convention 
involving $\Omega_a$. 
 
{\bf Lightlike ${\bm b^{\bm \mu}}$.} 
We select a coordinate system 
with $b^{\mu}=(B,\vec{B})$, 
where $|B|=|\vec{B}|$. 
In this frame, 
we obtain 
\beq{lightlike_roots} 
{}\hspace{-2mm}
(\lambda^{\pm}_a)^0= 
\pm\sqrt{m^2\hspace{-.6mm}+\hspace{-.6mm}\big[\vec{\lambda}\mp (-1)^a\vec{B}\,{\rm sgn}B\big]^2} \pm (-1)^a |B| 
\eeq 
for the roots of Eq.\ \rf{bdr}. 
Here, 
$a=1,2$ labels the $\Wb$ eigenvalue $\Omega_a$, 
which obeys the general relation \rf{genev}. 
To verify the correct labelling, 
we need to show 
that $\Omega_{a=1}$ ($\Omega_{a=2}$) is negative (positive). 
Using the result \rf{lightlike_roots} 
in Eq.\ \rf{bevfixed} yields 
$2\Omega_a=B^2\mp(-1)^a\,\vec{\lambda}\sdot\vec{B}\,{\rm sgn}B
+(-1)^a|B|\sqrt{m^2+\big[\vec{\lambda}\mp (-1)^a\vec{B}\,{\rm sgn}B\big]^2}$. 
If the square-root term 
dominates the two other terms on the right-hand side, 
its sign determines that of $\Omega_a$, 
and the correct choice of labels is confirmed. 
To see this,
we start with $(\vec{\lambda}\sdot\vec{B})^2<B^2\vec{\lambda}^2+m^2B^2$, 
which always holds for $m^2B^2\neq 0$. 
Adding $B^4\mp2(-1)^a B^2\,\vec{\lambda}\sdot\vec{B}\:{\rm sgn}B$ to both sides 
of this inequality 
and completing the squares gives 
$(B^2\mp(-1)^a\vec{\lambda}\cdot\vec{B}\,{\rm sgn}B)^2<
B^2(m^2+\big[\vec{\lambda}\mp (-1)^a\vec{B}\,{\rm sgn}B\big]^2)$, 
which establishes the claim. 

{\bf Spacelike ${\bm b^{\bm \mu}}$.} 
We choose our coordinate system such that $b^{\mu}=(0,\vec{B})$ 
with $|B|=|\vec{B}|$. 
We then find 
\beq{spacelike_roots} 
{}\hspace{-2.0mm}
(\lambda^{\pm}_a)^0\hspace{-.5mm}=
\pm\sqrt{\hspace{-.2mm}\vec{\lambda}^2\hspace{-.6mm}+\hspace{-.6mm}m^2\hspace{-.6mm}+\hspace{-.6mm}B^2\hspace{-.6mm}+
\hspace{-.6mm}2(-1)^a\sqrt{\hspace{-.2mm}m^2B^2\hspace{-.6mm}+\hspace{-.6mm}(\vec{\lambda}\sdot\vec{B})^2}}
\eeq
for the dispersion-relation solutions 
in this frame. 
It remains to verify 
that $a=1$ and $a=2$
lead to negative and positive $\Omega_a$, 
respectively. 
Employing the roots \rf{spacelike_roots} 
in Eq.\ \rf{bevfixed}, 
we obtain 
$2\Omega_a=(-1)^a\sqrt{\hspace{-.2mm}m^2B^2\hspace{-.6mm}+\hspace{-.6mm}(\vec{\lambda}\sdot\vec{B})^2}+B^2$. 
The square-root term dominates the right-hand side 
and therefore determines the sign of $\Omega_a$. 
This fact, which confirms the correct labeling, 
can be seen as follows. 
The lowest value $m|B|$ of the square root 
must be compared to the $B^2$ term. 
Since we are interested in small Lorentz violation $|b^2|\ll m^2$, 
we indeed have $B^2\ll mB$. 

{\bf Timelike ${\bm b^{\bm \mu}}$.} 
We will work in coordinates with $b^{\mu}=(B,\vec{0})$,  
which yields 
\beq{timelike_roots} 
(\lambda^{\pm}_a)^0=
\pm\sqrt{\left[|\vec{\lambda}|+(-1)^a|B|\right]^2+m^2}\;, 
\eeq
where again $a=1,2$. 
These dispersion-relation roots together with Eq.\ \rf{bevfixed} imply 
$2\Omega_a=(-1)^a|B||\vec{\lambda}|$ 
confirming consistent labeling.

\section{Invertibility of ${\bm\Wb}$}
\label{invert}

To study the invertibility of $\Wb$ on ${\cal S}_b$ and ${\cal S}_0$, 
we employ the eigenvalues \rf{genev} of this operator 
together with the appropriate dispersion relation. 
If the eigenvalues remain nonzero 
for all $(\lambda^\pm_a)^{\mu}(\vec{\lambda})$, 
$\Wb$ is invertible. 
Thus, 
\beq{invtest} 
(\lambda^\pm_a\!\cdot b)^2={\lambda^\pm_a}^2b^2
\eeq
needs to be investigated: 
if it is satisfied for some $\lambda_{\mu}(\vec{\lambda})$,  
$(\Wb)^{-1}$ is singular. 
In what follows, 
we separately consider 
the cases of lightlike, spacelike, and timelike $b^{\mu}$.

{\bf Lightlike ${\bm b^{\bm \mu}}$.} 
We choose a coordinate system with $b^{\mu}=(B,\vec{B})$, 
where $|B|=|\vec{B}|$. 
In the usual Dirac case, 
the formula \rf{invtest} gives 
$B^2[\pm(m^2+\vec{\lambda}^2)^{1/2}-|\vec{\lambda}|\cos\alpha]^2=0$, 
which cannot be satisfied for real $\vec{\lambda}$. 
In the nonzero $b^{\mu}$ case with roots \rf{lightlike_roots}, 
the requirement \rf{invtest} 
can be cast into the following form: 
$m^2+\vec{\lambda}^2\sin^2\alpha=0$, 
which again has no physical solutions. 
We conclude 
that for $b^2=0$, 
$\Wb$ is invertible 
on both ${\cal S}_b$ and ${\cal S}_0$. 

{\bf Spacelike ${\bm b^{\bm \mu}}$.} 
We select a coordinate system with $b^{\mu}=(0,\vec{B})$. 
For ordinary Dirac fermions, 
the singularity condition \rf{invtest} gives 
$(\lambda^\pm_a\cdot b)^2+\vec{B}^2m^2=0$, 
which cannot be satisfied for real quantities. 
We now turn our attention 
to the $b^\mu$ model with dispersion-relation roots \rf{spacelike_roots}. 
This equation 
together with the requirement \rf{invtest} leads to the relation 
$(\vec{\lambda}\cdot\vec{B})^2+\vec{B}^2(m^2-\vec{B}^2)=0$. 
On phenomenological grounds, 
$m^2\gg\vec{B}^2$ implying 
that physical solutions of this equation are impossible. 
We obtain the result 
that for $b^2<0$, 
$\Wb$ is invertible 
on ${\cal S}_b$ and ${\cal S}_0$. 

{\bf Timelike ${\bm b^{\bm \mu}}$.} 
We will work in coordinates with $b^{\mu}=(B,\vec{0})$. 
The singularity requirement \rf{invtest} 
takes then the simple form 
$\vec{\lambda}^2=0$ 
for arbitrary $(\lambda^\pm_a)^{\mu}$. 
This equation can be satisfied in 
both of the cases we are interested in. 
It follows 
that for $b^2>0$, 
the operator $\Wb$ fails to be invertible 
on the subspace spanned by 
plane waves of vanishing 3-momentum $\vec{\lambda}$. 
Note 
that $b^{\mu}$ and $(\lambda^\pm_a)^{\mu}$ are aligned 
in such cases, 
a result expected from the antisymmetric $\sigma^{\mu\nu}$ 
in the definition of $\Wb$.


\begin{thebibliography}{99}

\bibitem{cpt04}
V.A.\ Kosteleck\'y, ed.,
{\it CPT and Lorentz Symmetry III},
World Scientific, Singapore, 2005;
{\it CPT and Lorentz Symmetry II},
World Scientific, Singapore, 2002;
{\it CPT and Lorentz Symmetry},
World Scientific, Singapore, 1999.

\bibitem{reviews}  
R.\ Bluhm, 
arXiv:hep-ph/0506054; 
D.\ Mattingly, 
Living Rev.\ Rel.\  {\bf 8}, 5 (2005) 
[arXiv:gr-qc/0502097]; 
G.\ Amelino-Camelia \etal, 
AIP Conf.\ Proc.\ {\bf 758}, 30 (2005) 
[arXiv:gr-qc/0501053]; 

\bibitem{ck97}
D.\ Colladay and V.A.\ Kosteleck\'y, 
Phys.\ Rev.\ D {\bf 55}, 6760 (1997)
[arXiv:hep-ph/9703464].

\bibitem{sme} 
D.\ Colladay and V.A.\ Kosteleck\'y, 
Phys.\ Rev.\ D {\bf 58}, 116002 (1998) 
[arXiv:hep-ph/9809521]. 

\bibitem{gravext} 
V.A.\ Kosteleck\'y, 
Phys.\ Rev.\ D {\bf 69}, 105009 (2004) 
[arXiv:hep-th/0312310]. 

\bibitem{photonexpt}
S.M.\ Carroll, G.B.\ Field, and R.\ Jackiw, 
Phys.\ Rev.\ D {\bf 41}, 1231 (1990); 
J.\ Lipa \etal, 
Phys.\ Rev.\ Lett.\  {\bf 90}, 060403 (2003) 
[arXiv:physics/0302093]; 
H.\ M\"uller \etal, 
Phys.\ Rev.\ Lett.\ {\bf 91}, 020401 (2003) 
[arXiv:physics/0305117]; 
P.\ Wolf \etal, 
Gen.\ Rel.\ Grav.\ {\bf 36}, 2352 (2004) 
[arXiv:gr-qc/0401017]; 
Phys.\ Rev.\ D {\bf 70}, 051902 (2004) 
[arXiv:hep-ph/0407232]; 
M.E.\ Tobar \etal, 
Phys.\ Rev.\ D {\bf 71}, 025004 (2005) 
[arXiv:hep-ph/0408006]; 
P.L.\ Stanwix \etal, 
arXiv:gr-qc/0609072; 
S.\ Herrmann \etal, 
Phys.\ Rev.\ Lett.\  {\bf 95}, 150401 (2005) 
[arXiv:physics/0508097]; 
V.A.\ Kosteleck\'y and M.\ Mewes, 
Phys.\ Rev.\ Lett.\ {\bf 87}, 251304 (2001) 
[arXiv:hep-ph/0111026]; 
Phys.\ Rev.\ Lett.\ (in press) 
[arXiv:hep-ph/0607084]. 

\bibitem{radiative}
R.\ Jackiw and V.A.\ Kosteleck\'y, 
Phys.\ Rev.\ Lett.\ {\bf 82}, 3572 (1999)
[arXiv:hep-ph/9901358];
M.\ P\'erez-Victoria,
Phys.\ Rev.\ Lett.\  {\bf 83}, 2518 (1999)
[arXiv:hep-th/9905061]; 
V.A.\ Kosteleck\'y and A.G.M.\ Pickering, 
Phys.\ Rev.\ Lett.\  {\bf 91}, 031801 (2003) 
[arXiv:hep-ph/0212382]; 
B.\ Altschul,
Phys.\ Rev.\ D {\bf 70}, 101701 (2004)
[arXiv:hep-th/0407172].

\bibitem{photonth2} 
C.\ Adam and F.R.\ Klinkhamer, 
Nucl.\ Phys.\ B {\bf 657}, 214 (2003) 
[arXiv:hep-th/0212028]; 
T.\ Jacobson, S.\ Liberati, and D.\ Mattingly, 
Phys.\ Rev.\ D {\bf 67}, 124011 (2003) 
[arXiv:hep-ph/0209264]; 
V.A.\ Kosteleck\'y and A.G.M.\ Pickering, 
Phys.\ Rev.\ Lett.\ {\bf 91}, 031801 (2003) 
[arXiv:hep-ph/0212382]; 
R.\ Lehnert, 
Phys.\ Rev.\ D {\bf 68}, 085003 (2003) 
[arXiv:gr-qc/0304013]; 
R.\ Lehnert and R.\ Potting, 
Phys.\ Rev.\ Lett.\ {\bf 93}, 110402 (2004) 
[arXiv:hep-ph/0406128]; 
Phys.\ Rev.\ D {\bf 70}, 125010 (2004) 
[arXiv:hep-ph/0408285]; 
F.R.\ Klinkhamer and C.\ Rupp, 
Phys.\ Rev.\ D {\bf 72}, 017901 (2005) 
[arXiv:hep-ph/0506071]; 
C.\ Kaufhold and F.R.\ Klinkhamer, 
Nucl.\ Phys.\ B {\bf 734}, 1 (2006) 
[arXiv:hep-th/0508074]; 
B.\ Altschul, 
Phys.\ Rev.\ D {\bf 72}, 085003 (2005) 
[arXiv:hep-th/0507258]; 
Phys.\ Rev.\ Lett.\ {\bf 96}, 201101 (2006) 
[arXiv:hep-ph/0603138]; 
arXiv:hep-th/0609030; 
B.\ Feng \etal, 
Phys.\ Rev.\ Lett.\  {\bf 96}, 221302 (2006) 
[arXiv:astro-ph/0601095]; 
C.D.\ Carone, M.\ Sher, and M.\ Vanderhaeghen, 
arXiv:hep-ph/0609150. 

\bibitem{km02} 
V.A.\ Kosteleck\'y and M.~Mewes, 
Phys.\ Rev.\ D {\bf 66}, 056005 (2002) 
[arXiv:hep-ph/0205211]. 

\bibitem{trap} 
H.\ Dehmelt \etal, 
Phys.\ Rev.\ Lett.\  {\bf 83}, 4694 (1999) 
[arXiv:hep-ph/9906262]; 
R.\ Mittleman \etal, 
Phys.\ Rev.\ Lett.\ {\bf 83}, 2116 (1999); 
G.\ Gabrielse \etal, 
Phys.\ Rev.\ Lett.\ {\bf 82}, 3198 (1999); 
R.\ Bluhm \etal, 
Phys.\ Rev.\ Lett.\ {\bf 79}, 1432 (1997) 
[arXiv:hep-ph/9707364]; 
Phys.\ Rev.\ D {\bf 57}, 3932 (1998) 
[arXiv:hep-ph/9809543]; 
Phys.\ Rev.\ Lett.\ {\bf 82}, 2254 (1999) 
[arXiv:hep-ph/9810269]. 

\bibitem{spinpol} 
L.-S.\ Hou, W.-T.\ Ni, and Y.-C.M.\ Li, 
Phys.\ Rev.\ Lett.\ {\bf 90}, 201101 (2003) 
[arXiv:physics/0009012]; 
R.\ Bluhm and V.A.\ Kosteleck\'y, 
Phys.\ Rev.\ Lett.\  {\bf 84}, 1381 (2000) 
[arXiv:hep-ph/9912542]; 
B.R.\ Heckel \etal, 
Phys.\ Rev.\ Lett.\  {\bf 97}, 021603 (2006) 
[arXiv:hep-ph/0606218]. 

\bibitem{eexpt3} 
V.A.\ Kosteleck\'y, C.D.\ Lane and A.G.M.\ Pickering, 
Phys.\ Rev.\ D {\bf 65}, 056006 (2002) 
[arXiv:hep-th/0111123]; 
H.\ M\"uller \etal, 
Phys.\ Rev.\ D {\bf 68}, 116006 (2003) 
[arXiv:hep-ph/0401016]; 
Phys.\ Rev.\ D {\bf 70}, 076004 (2004) 
[arXiv:hep-ph/0405177]; 
H.\ M\"uller, 
Phys.\ Rev.\ D {\bf 71}, 045004 (2005) 
[arXiv:hep-ph/0412385]. 

\bibitem{cc} 
D.\ Bear \etal, 
Phys.\ Rev.\ Lett.\ {\bf 85}, 5038 (2000) 
[arXiv:physics/0007049]; D.F.\ Phillips \etal, 
Phys.\ Rev.\ D {\bf 63}, 111101 (2001) 
[arXiv:physics/0008230]; 
M.A.\ Humphrey \etal, 
Phys.\ Rev.\ A {\bf 62}, 063405 (2000) 
[arXiv:physics/0007056]; 
Phys.\ Rev.\ A {\bf 68}, 063807 (2003); 
F.\ Can\`e \etal, 
Phys.\ Rev.\ Lett.\ {\bf 93}, 230801 (2004) 
[arXiv:physics/0309070]; 
P.\ Wolf \etal, 
Phys.\ Rev.\ Lett.\ {\bf 96}, 060801 (2006) 
[arXiv:hep-ph/0601024]; 
V.A.\ Kosteleck\'y and C.D.\ Lane, 
Phys.\ Rev.\ D {\bf 60}, 116010 (1999) 
[arXiv:hep-ph/9908504]; 
C.D.\ Lane,  
Phys.\ Rev.\ D {\bf 72}, 016005 (2005) 
[arXiv:hep-ph/0505130]. 

\bibitem{spaceexpt} 
R.\ Bluhm \etal, 
Phys.\ Rev.\ Lett.\ {\bf 88}, 090801 (2002) 
[arXiv:hep-ph/0111141]; 
Phys.\ Rev.\ D {\bf 68}, 125008 (2003) 
[arXiv:hep-ph/0306190]. 

\bibitem{bnsyn} 
O.\ Bertolami \etal, 
Phys.\ Lett.\ B {\bf 395}, 178 (1997) 
[arXiv:hep-ph/9612437]; 
G.\ Lambiase, 
Phys.\ Rev.\ D {\bf 72}, 087702 (2005) 
[arXiv:astro-ph/0510386]; 
J.M.\ Carmona \etal, 
Mod.\ Phys.\ Lett.\ A {\bf 21}, 883 (2006) 
[arXiv:hep-th/0410143]; 
S.M.\ Carroll and J.\ Shu, 
Phys.\ Rev.\ D {\bf 73}, 103515 (2006) 
[arXiv:hep-ph/0510081]. 

\bibitem{hadronexpt}
KTeV Collaboration, 
H.\ Nguyen, 
in 
{\it CPT and Lorentz Symmetry II}, Ref.\ \cite{cpt04}; 
OPAL Collaboration, 
K.\ Ackerstaff \etal, 
Z.\ Phys.\ C {\bf 76}, 401 (1997) 
[arXiv:hep-ex/9707009]; 
DELPHI Collaboration, 
M.\ Feindt \etal, 
preprint DELPHI 97-98 CONF 80 (1997); 
BELLE Collaboration, 
K.\ Abe \etal, 
Phys.\ Rev.\ Lett.\ {\bf 86}, 3228 (2001) 
[arXiv:hep-ex/0011090]; 
BaBar Collaboration, 
B.\ Aubert \etal, 
Phys.\ Rev.\ Lett.\ {\bf 92}, 181801 (2004) 
[arXiv:hep-ex/0311037]; 
arXiv:hep-ex/0607103; 
FOCUS Collaboration, 
J.M.\ Link \etal, 
Phys.\ Lett.\ B {\bf 556}, 7 (2003) 
[arXiv:hep-ex/0208034]; 
V.A.\ Kosteleck\'y and R.\ Potting, 
Phys.\ Rev.\ D {\bf 51}, 3923 (1995) 
[arXiv:hep-ph/9501341]; 
V.A.\ Kosteleck\'y, 
Phys.\ Rev.\ Lett.\ {\bf 80}, 1818 (1998) 
[arXiv:hep-ph/9809572]; 
Phys.\ Rev.\ D {\bf 61}, 016002 (2000) 
[arXiv:hep-ph/9909554]. 

\bibitem{muexpt} 
V.W.\ Hughes \etal, 
Phys.\ Rev.\ Lett.\ {\bf 87}, 111804 (2001) 
[arXiv:hep-ex/0106103]; 
R.\ Bluhm \etal, 
Phys.\ Rev.\ Lett.\ {\bf 84}, 1098 (2000) 
[arXiv:hep-ph/9912451]. 

\bibitem{neutrinos} 
V.D.\ Barger \etal, 
Phys.\ Rev.\ Lett.\  {\bf 85}, 5055 (2000) 
[arXiv:hep-ph/0005197]; 
A.\ de Gouv\^ea, 
Phys.\ Rev.\ D {\bf 66}, 076005 (2002) 
[arXiv:hep-ph/0204077]. 

\bibitem{nuexpt} 
LSND Collaboration, 
L.B.\ Auerbach \etal, 
Phys.\ Rev.\ D {\bf 72}, 076004 (2005) 
[arXiv:hep-ex/0506067]; 
M.D.\ Messier (SK), 
in {\it CPT and Lorentz Symmetry III}, 
Ref.\ \cite{cpt04}; 
B.J.\ Rebel and S.F.\ Mufson (MINOS), 
in {\it CPT and Lorentz Symmetry III}, 
Ref.\ \cite{cpt04}; 
V.A.\ Kosteleck\'y and M.\ Mewes, 
Phys.\ Rev.\ D {\bf 69}, 016005 (2004) 
[arXiv:hep-ph/0309025]; 
Phys.\ Rev.\ D {\bf 70}, 031902(R) (2004) 
[arXiv:hep-ph/0308300]; 
Phys.\ Rev.\ D {\bf 70}, 076002 (2004) 
[arXiv:hep-ph/0406255]; 
T.\ Katori \etal, 
arXiv:hep-ph/0606154. 

\bibitem{higgs} 
D.L.\ Anderson, M.\ Sher, and I.\ Turan, 
Phys.\ Rev.\ D {\bf 70}, 016001 (2004) 
[arXiv:hep-ph/0403116]; 
E.O.\ Iltan, 
Mod.\ Phys.\ Lett.\ A {\bf 19}, 327 (2004) 
[arXiv:hep-ph/0309154]. 

\bibitem{bk06}
Q.G.\ Bailey and V.A.\ Kosteleck\'y,
Phys.\ Rev.\ D {\bf 74}, 045001 (2006) 
[arXiv:gr-qc/0603030]. 

\bibitem{shore} 
For a related work, 
see, e.g., 
G.M.\ Shore, 
Nucl.\ Phys.\ B {\bf 717}, 86 (2005) 
[arXiv:hep-th/0409125]. 

\bibitem{ksp} 
V.A.\ Kosteleck\'y and S.\ Samuel, 
Phys.\ Rev.\ D {\bf 39}, 683 (1989); 
V.A.\ Kosteleck\'y and R.\ Potting, 
Nucl.\ Phys.\ B {\bf 359}, 545 (1991). 
R.\ Bluhm and V.A.\ Kosteleck\'y, 
Phys.\ Rev.\ D {\bf 71}, 065008 (2005) 
[arXiv:hep-th/0412320] 
B.\ Altschul and V.A.\ Kosteleck\'y, 
Phys.\ Lett.\ B {\bf 628}, 106 (2005) 
[arXiv:hep-th/0509068]; 
V.A.\ Kosteleck\'y and R.\ Potting, 
Gen.\ Rel.\ Grav.\  {\bf 37}, 1675 (2005) 
[arXiv:gr-qc/0510124]. 

\bibitem{gyb} 
G.Yu.\ Bogoslovsky, 
arXiv:math-ph/0511077; 
Phys.\ Lett.\ A {\bf 350}, 5 (2006) 
[arXiv:hep-th/0511151]. 

\bibitem{ncqed} 
See, e.g., 
I.\ Mocioiu, M.\ Pospelov, and R.\ Roiban, 
Phys.\ Lett.\ B {\bf 489}, 390 (2000) 
[arXiv:hep-ph/0005191]; 
S.M.\ Carroll \etal, 
Phys.\ Rev.\ Lett.\ {\bf 87}, 141601 (2001) 
[arXiv:hep-th/0105082]; 
Z.\ Guralnik, R.\ Jackiw, S.Y.\ Pi, and A.P.\ Polychronakos, 
Phys.\ Lett.\ B {\bf 517}, 450 (2001) 
[arXiv:hep-th/0106044]; 
C.E.\ Carlson, C.D.\ Carone, and R.F.\ Lebed, 
Phys.\ Lett.\ B {\bf 518}, 201 (2001) 
[arXiv:hep-ph/0107291]; 
A.\ Anisimov \etal, 
Phys.\ Rev.\ D {\bf 65}, 085032 (2002) 
[arXiv:hep-ph/0106356]; 
A.\ Das \etal, 
Phys.\ Rev.\ D {\bf 72}, 107702 (2005) 
[arXiv:hep-th/0510002]. 

\bibitem{spacetimevarying} 
V.A.\ Kosteleck\'y, R.\ Lehnert, and M.\ Perry, 
Phys.\ Rev.\ D {\bf 68}, 123511 (2003) 
[arXiv:astro-ph/0212003]; 
R.\ Jackiw and S.-Y.\ Pi, 
Phys.\ Rev.\ D {\bf 68}, 104012 (2003) 
[arXiv:gr-qc/0308071]; 
O.\ Bertolami \etal, 
Phys.\ Rev.\ D {\bf 69}, 083513 (2004) 
[arXiv:astro-ph/0310344]; 

\bibitem{ghostcond} 
N.\ Arkani-Hamed \etal, 
JHEP {\bf 0507}, 029 (2005)
[arXiv:hep-ph/0407034].

\bibitem{qg}
J.\ Alfaro, H.A.\ Morales-T\'ecotl, and L.F.\ Urrutia, 
Phys.\ Rev.\ D {\bf 66}, 124006 (2002) 
[arXiv:hep-th/0208192]; 
D.\ Sudarsky, L.\ Urrutia, and H.\ Vucetich, 
Phys.\ Rev.\ Lett.\ {\bf 89}, 231301 (2002) 
[arXiv:gr-qc/0204027]; 
Phys.\ Rev.\ D {\bf 68}, 024010 (2003) 
[arXiv:gr-qc/0211101]; 
G.\ Amelino-Camelia, 
Mod.\ Phys.\ Lett.\ A {\bf 17}, 899 (2002) 
[arXiv:gr-qc/0204051]; 
R.\ Myers and M.\ Pospelov, 
Phys.\ Rev.\ Lett.\ {\bf 90}, 211601 (2003) 
[arXiv:hep-ph/0301124]; 
N.E.\ Mavromatos, 
Lect.\ Notes Phys.\ {\bf 669}, 245 (2005) 
[arXiv:gr-qc/0407005]. 

\bibitem{klink} 
F.R.\ Klinkhamer and J.\ Schimmel, 
Nucl.\ Phys.\ B {\bf 639}, 241 (2002) 
[arXiv:hep-th/0205038]; 
F.R.\ Klinkhamer and C.\ Rupp, 
Phys.\ Rev.\ D {\bf 70}, 045020 (2004) 
[arXiv:hep-th/0312032]. 

\bibitem{fn02} 
C.D.\ Froggatt and H.B.\ Nielsen, 
arXiv:hep-ph/0211106. 

\bibitem{bj} 
J.D.\ Bjorken, 
Phys.\ Rev.\ D {\bf 67}, 043508 (2003) 
[arXiv:hep-th/0210202]. 

\bibitem{brane} 
C.P.\ Burgess \etal, 
JHEP {\bf 0203}, 043 (2002) 
[arXiv:hep-ph/0201082]; 
A.R.\ Frey, 
JHEP {\bf 0304}, 012 (2003) 
[arXiv:hep-th/0301189]; 
J.\ Cline and L.\ Valc\'arcel, 
JHEP {\bf 0403}, 032 (2004) 
[arXiv:hep-ph/0312245]. 

\bibitem{redef} 
M.S.\ Berger and V.A.\ Kosteleck\'y, 
Phys.\ Rev.\ D {\bf 65}, 091701 (2002) 
[arXiv:hep-th/0112243]. 
D.\ Colladay and P.\ McDonald, 
J.\ Math.\ Phys.\  {\bf 43}, 3554 (2002) 
[arXiv:hep-ph/0202066]; 
B.\ Altschul,
arXiv:hep-th/0602235. 

\bibitem{kl01} 
V.A.\ Kosteleck\'y and R.\ Lehnert,
Phys.\ Rev.\ D {\bf 63}, 065008 (2001)
[arXiv:hep-th/0012060].

\bibitem{finiteT} 
L.\ Cervi, L.\ Griguolo, and D.\ Seminara, 
Phys.\ Rev.\ D {\bf 64}, 105003 (2001) 
[arXiv:hep-th/0104022]; 
T.\ Mariz \etal, 
JHEP {\bf 0510}, 019 (2005) 
[arXiv:hep-th/0509008]. 

\bibitem{decay}
V.C.\ Zhukovsky, A.E.\ Lobanov, and E.M.\ Murchikova,
Phys.\ Rev.\ D {\bf 73}, 065016 (2006)
[arXiv:hep-ph/0510391].

\bibitem{IZ} 
See, e.g., 
C.\ Itzykson and J.-B.\ Zuber, 
{\it Quantum Field Theory}, 
(McGraw--Hill, Singapore, 1980). 

\bibitem{rl04} 
R.\ Lehnert, 
J.\ Math.\ Phys.\  {\bf 45}, 3399 (2004) 
[arXiv:hep-ph/0401084]. 

\bibitem{ck01} 
D.\ Colladay and V.A.\ Kosteleck\'y, 
Phys.\ Lett.\ B {\bf 511}, 209 (2001) 
[arXiv:hep-ph/0104300]. 

\end{thebibliography}
\end{document}